\documentclass[fleqn,10pt]{wlscirep}
\usepackage{hyperref}
\usepackage[utf8]{inputenc}
\usepackage[T1]{fontenc}
\usepackage{booktabs}
\usepackage{multirow}
\usepackage{verbatim}
\usepackage{subcaption}
\usepackage{comment}
\usepackage{amsmath,amssymb}

\newcommand{\smallotimes}

\title{Deep-learning-aided dismantling of interdependent networks}

\author[1]{Weiwei Gu}
\author[1,*]{Chen Yang}
\author[2]{Lei Li}
\author[1]{Jinqiang Hou}
\author[3,*]{Filippo Radicchi}
\affil[1]{UrbanNet Lab, College of Information Science and Technology, Beijing University of Chemical Technology, Beijing, 100029, P. R. China}
\affil[2]{School of Software Enginering, University of Science and Technology of China, Hefei, 230026, P. R. China}
\affil[3]{Center for Complex Networks and Systems Research, Luddy School of Informatics, Computing, and Engineering, Indiana University, Bloomington, Indiana 47408, USA}
\affil[*]{To whom correspondence should be addressed; E-mail: chenyang@buct.edu.cn, filiradi@iu.edu}

\begin{abstract}
Identifying the minimal set of nodes whose removal 
breaks a complex network apart, also referred as the network dismantling problem, 
is a highly non-trivial task with applications
in multiple domains.
Whereas network dismantling has been extensively studied over the past decade, research has primarily focused on the formulations of the optimization problem for single-layer networks, neglecting that many, if not all, real networks display multiple layers of interdependent interactions.
In such networks, the optimization problem is fundamentally different as the effect of removing nodes propagates within and across layers in a way that can not be predicted using a single-layer perspective.
Here, we propose a dismantling algorithm named MultiDismantler, which leverages multiplex network representation and deep reinforcement learning to 
optimally dismantle multi-layer interdependent networks.
MultiDismantler is trained on small synthetic multiplex graphs; when applied to large, real and synthetic networks, it displays 
exceptional dismantling performance, clearly outperforming
all existing methods that rely on a single-layer 
approach to network dismantling. 
We show that MultiDismantler is effective in guiding strategies for the containment of diseases in social networks
characterized by multiple layers of social interactions. Also, we show that MultiDismantler is useful in the design of protocols aimed at delaying the onset of cascading failures in interdependent critical infrastructures. 

\end{abstract}

\begin{document}

\flushbottom

\maketitle
\par\smallskip
\noindent\begin{minipage}{\textwidth}\small
\textbf{Author Accepted Manuscript.} Published as W. Gu, C. Yang, L. Li, J. Hou, and F. Radicchi, ``Deep-learning-aided dismantling of interdependent networks,'' \textit{Nature Machine Intelligence} \textbf{7}, 1266--1277 (2025). DOI: \href{https://doi.org/10.1038/s42256-025-01070-2}{10.1038/s42256-025-01070-2}.
\end{minipage}\par\medskip

\thispagestyle{empty}

\section{Introduction}

Networks are ubiquitous data structures used to model various complex systems, including natural systems, e.g., climate networks and protein-protein interaction networks, and man-made systems, e.g., social networks and infrastructural networks~\cite{watts1998collective,boccaletti2006complex}. Connectedness is essential for the overall network function, as being part of the same connected component is a necessary condition for interaction~\cite{cohen2001breakdown}. Due to their heterogeneous structure, connectedness in real networks hinges on specific sets of nodes. Identifying the minimal set of nodes whose removal would destroy any extensively connected components in a network is known as the network dismantling problem~\cite{morone2015influence}. 
Network dismantling has numerous practical applications. For example, 
optimally inhibiting specific enzymes or proteins for drug design
is essentially a network dismantling problem~\cite{barabasi2011network}. Similarly, enhancing the tolerance and robustness of a small set of nodes, such as traffic or electronic sites, to maximally improve the effectiveness and efficiency of infrastructural networks can be thought as network dismantling. ~\cite{carreras2002critical,bertagnolli2021quantifying}. Finally,
finding the smallest set of people that should be vaccinated to prevent a disease outbreak is well approximated by the solution of a network dismantling problem~\cite{morone2015influence,chen2008finding}.

Network dismantling is an NP-hard (non-deterministic polynomial time) problem and
is among the most fundamental optimization problems that can be formulated on a network~\cite{braunstein2016network, artime2024robustness}.
Due to the complexity of the optimization problem, 
exact solutions of the network dismantling problem can be
obtained on small networks only. In medium and large networks, such as those typically considered in practical applications of the network dismantling problem, solutions can only be approximated.
Many existing dismantling algorithms provide approximate solutions
that are based on node-centrality heuristics, such as
adaptive High Degree (HDA)\cite{chen2009efficient}, Collective Influence (CI)\cite{morone2015influence}, and MIN-SUM\cite{braunstein2016network}. These heuristic methods do not optimize a global function, provide no performance guarantees, and lack generalization capabilities. Machine learning-based attack strategies such as FINDER\cite{fan2020finding}, GDM\cite{grassia2021machine}, and NIRM\cite{zhang2022dismantling} transform network dismantling into a discrete combinatorial optimization problem, learning dismantling policies from exact solutions on small synthetic Barabási-Albert (BA) networks; these policies appear successful when used to approximate solutions of large-scale network dismantling problems. 

The aforementioned algorithms focus solely on single, isolated networks, neglecting the fact that real networks are often formed by multiple interdependent layers~\cite{bianconi2018multilayer}.
A network composed of interdependent layers is generally more vulnerable than each of its network layers considered in isolation. Localized damage of the nodes in one network layer may lead to the failure of the nodes in the other layers, causing cascading failures throughout the entire system~\cite{buldyrev2010catastrophic}. This cascading mechanism makes system collapse more difficult to anticipate and control\cite{vespignani2010fragility}. Examples include the 2003 blackout in Italy~\cite{buldyrev2010catastrophic} and the sudden disruptions of the global supply chain during the COVID-19 pandemic in 2020~\cite{chen2023epidemic}. Identifying the minimal set of nodes whose removal leads a multi-layer network to collapse is crucial for 
all these real systems, however, the dismantling problem of a network composed of interdependent layers can not simply be solved by considering each of its layers in isolation~\cite{osat2017optimal, baxter2018targeted, coghi2018controlling}.

Motivated by recent advancements in applying machine learning techniques to graph optimization problems~\cite{schuetz2022combinatorial,khalil2017learning}, we propose here a deep-learning-aided dismantling framework named MultiDismantler, which can approximate solutions in large-scale multi-layer networks with high effectiveness. MultiDismantler first generates a large number of small synthetic networks using the Geometric Multiplex Model (GMM)~\cite{kleineberg2016hidden}. This approach overcomes the limitations of the BA model which cannot characterize interactions and correlations across different network layers. Geometric correlations among network layers are ubiquitously present in real multi-layer networks~\cite{kleineberg2016hidden,boguna2021network}; those correlations are known to dramatically affect the outcome of both structural and dynamical processes occurring on the networks~\cite{kleineberg2017geometric,faqeeh2018characterizing,patwardhan2023epidemic}. After training corpus generation, MultiDismantler designs a graph convolutional neural network model with a flexible node-level inter-layer attention mechanism to encode nodes and states in multiplex networks. This is followed by an n-step Deep Q-Network (DQN)~\cite{sutton1988learning}, which takes node states as input and aims to automatically learn an optimization strategy that quickly dismantles multi-layer interdependent networks.


MultiDismantler, trained on small synthetic multiplex graphs, displays 
excellent 
performance across various complex real-world multiplex networks, handling 
arbitrary removal node costs~\cite{ren2019generalized}. MultiDismantler is characterized by outstanding generalization capabilities. 
Our findings indicate that the node set identified by MultiDismantler aligns with the optimal set obtained via the exhaustive brute-force method, where all possible node sequence combinations are tested to find the optimal set that fully dismantles the network. This consistency not only validates the effectiveness of MultiDismantler but also provides evidence of the potential of deep reinforcement learning techniques in generating effective solutions for NP-hard graph optimization problems.


We validate MultiDismantler in two practical applications of network dismantling. First, we use MultiDismantler to design optimal immunization strategies in spreading processes on multi-layer social networks; second, we take advantage of MultiDismantler as a tool to mitigate system collapse in interdependent networks. In both applications, MultiDismantler outperforms other dismantling algorithms adapted from single- to multi-layer networks.


\section{Results}

\subsection{Dismantling interdependent networks}
We consider a network 
consisting of  two layers of interactions. Nodes in the two layers  are one-to-one
interdependent. For simplicity of description but without loss of generality, we do not make a formal distinction between one node and its replica in the other layer, rather we act as the two layers would share the same set of nodes. Depending on the context, networks of this type, i.e., with distinguishable layers of interaction but indistinguishable nodes across the layers, are also referred as multiplex networks or edge-colored graphs~\cite{ramsey1987problem,kivela2014multilayer}. 
Two nodes in the network are in the same mutually connected component if they are connected through edges that involve nodes in the very same mutually connected component in each of the layers of the network~\cite{buldyrev2010catastrophic}. The Largest Mutually Connected Component (LMCC) is the largest of such mutually connected components. 

The network dismantling problem involves identifying the optimal sequence of nodes whose removal efficiently results in the fragmentation of the network into numerous disconnected components of non-extensive size~\cite{osat2017optimal} (see Section 4.1.1 for the definition of the optimization problem). The removal of each node in the sequence incurs a cost~\cite{ren2019generalized}. We consider two scenarios: i) the cost of each node is the same, i.e., unit cost, and ii) the cost of a node equals the number of edges that are removed together with the node, i.e., degree cost. Also, we relax scenario (i) by allowing nodes to have cost values taken at random from the uniform, normal, and Poisson distributions. This scenario is used to test the generalizability of our algorithm to cases where removal costs are uncertain. The size of the LMCC provides a natural metric to monitor the fragmentation of the network into non-extensive mutually connected components; such a quantity is monitored as a function of the cost of the sequence of nodes removed from the network. We note that the size of the LMCC is a decreasing function of the number of nodes removed from the network. The Area Under the Dismantling Curve (AUDC), i.e., the relative size of the LMCC as a function of the relative cost of the removed nodes, is a natural quantity to assess the quality of a given sequence~\cite{fan2020finding,grassia2021machine}. Another natural metric to assess the quality of a specific sequence of nodes provided by a dismantling algorithm is given by the cost required to make the LMCC non extensive. The dismantling cost, namely $C^*$, is given by the smallest value of the cost required to make the LMCC smaller than this threshold value. We normalize such a dismantling cost by dividing it by overall cost of removing all nodes in the network. Also, we adopt here the standard convention of defining the LMCC as non extensive if its size is smaller than the square root of the original LMCC~\cite{clusella2016immunization}.

\subsection{Proposed approach to the dismantling of interdependent networks}
MultiDismantler is the first approach able to integrate multiplex network representation learning with a deep reinforcement learning architecture to solve multiplex optimization problems. To ensure success, three main challenges are addressed
by MultiDismantler:

\begin{enumerate}

\item MultiDismantler is trained on synthetic graphs
constructed via a generative network model able to accurately  
reproduce the structure of real-world interdependent networks.
This approach is essential for mimicking the diversity of real-world networks with synthetic networks, thereby increasing the generalizability of MultiDismantler.

\item MultiDismantler relies on an encoding framework designed to effectively capture and preserve both intra- and inter-layer network structures and properties. This ensures high-quality representations that facilitate learning and decision-making.

\item MultiDismantler appropriately bridges 
multi-layer representation learning vectors with deep reinforcement learning components. This involves defining the network state, specifying node deletion actions, designing the environment, and establishing evaluation metrics to guide the learning process.

\end{enumerate}

\begin{figure}[htbp]  
\centering  
\includegraphics[width=\linewidth]{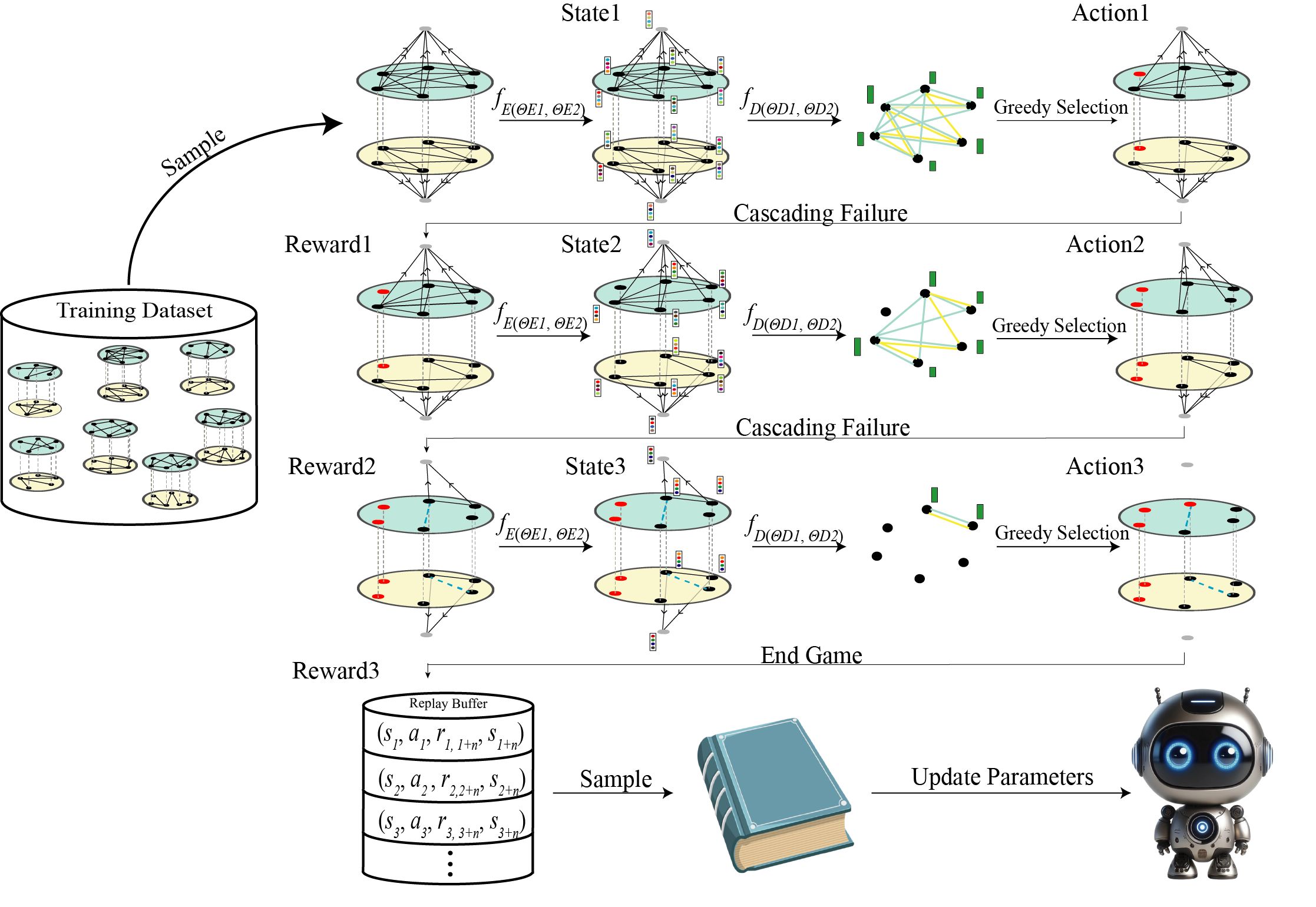}  
\caption{The training process of MultiDismantler. The process includes generating training networks, dismantling such networks, and learning from previous experience. The training data includes over 20,000 synthetic interdependent networks, with size from 30 to 50 nodes, generated according to the Geometric Multiplex Model. 
To disintegrate a network, we first add a virtual node (denoted in grey) to each layer; all nodes in the layer are neighbors of the virtual nodes, but not vice versa.
With the Multiplex Graph Neural Network (MGNN) model which consists of a graph convolutional neural network model( $\Theta_{E1}$) and
node-level inter-layer attention mechanism ($\Theta_{E2}$),
MultiDismantler encodes multiplex networks, including the embedding of virtual nodes as well as the final embedding of each node. Subsequently, the agent decodes the embedding vectors using multi-layer perceptrons (MLPs) ($\Theta_{D1}$) and layer-level attention mechanism ($\Theta_{D2}$) to compute the expected return (Q-value) of each node, which reflects the expected reward that can be obtained by removing a node under the current network conditions. The agent employs an epsilon-greedy strategy to remove nodes (red nodes): with probability 1-$\epsilon$, it selects the node with the highest Q-value, otherwise, it randomly chooses the removal node. Considering the interdependencies between layers in a multiplex network, a cascading failure may occur after the agent removes a node, leading to the removal of more edges (blue lines). After the node deletion operation, the agent re-encodes and decodes the residual network, ranking the importance of each node, until the network is completely dismantled (LMCC=1). 
Once the network is fully dismantled, we collect the n-step information in the form of $( s_{t}, a_{t}, r_{t, t+n}, s_{t+n} ) $  tuples and store them in an experience replay buffer. The agent learns by randomly sampling mini-batches from the experience replay buffer to update its parameters.}  
\label{fig:framework}  
\end{figure}

A schematic illustration of the MultiDismantler framework is provided in Fig.~\ref{fig:framework}. Below, we briefly describe the three main components of 
MultiDismantler.

\textbf{Geometric Multiplex Model (GMM)}. GMM generates synthetic multiplex networks with cross-layer geometric correlations. 
GMM has been widely used for understanding behaviors of real multi-layer networks\cite{kleineberg2016hidden,kleineberg2017geometric}. The model has several tunable parameters such as the inter-layer degree correlation $d$ and the inter-layer similarity correlation $g$; at the individual layer level, the tunable parameters are the power-law degree distribution exponent $\gamma$, the expected mean degree for each layer $\overline{k}$ and the network clustering control parameter $T$. The inter-layer similarity $g$ plays a vital role in network disintegration prediction\cite{kleineberg2017geometric}. In this paper, we consider different $g$ values, and find that the best dismantling performance is obtained by setting $g = 0.5$ (SI 4C). 

\textbf{Multiplex Representation Learning.} Traditional representation learning frameworks
for multi-layer networks, often utilizing graph neural network architectures, typically employ meta-paths or layer-level attention mechanisms to encode nodes across layers~\cite{ma2019multi,dong2017metapath2vec,xu2021topic,gu2024pay}. 
These frameworks independently learn node representations in each layer and concatenate them using deep learning techniques such as multi-layer perceptions; however, they fail to fully capture the complexity of the inter-layer topology.
In this paper, we propose a new Multiplex Graph Neural Network (MGNN) model, which employs a series of layer-specific graph convolution networks to obtain intra-layer representations by aggregating information from neighboring nodes during the message propagation process (see Eq.~\ref{eq:intra_aggregate}). After intra-layer propagation, MGNN applies a multi-faceted linear transformation matrix to project intra-layer representations into a shared representation space (see Eq.~\ref{eq:mapping}). Incorporating feature influence from cross-layer neighbors can enrich node representations; however, the impact of inter-layer neighbors on different nodes varies significantly. Some nodes are influenced primarily by intra-layer connections, while others rely more on inter-layer edge information. MGNN incorporates a node-level attention mechanism that computes customized attention weights for each node (see Eq.~\ref{eq:bitwise}). The final representation of a node in a given layer is a weighted combination of its neighbors' representations from both intra-layer and inter-layer connections (see Eq.~\ref{eq:inter_aggregate}).

\textbf{Dismantling Strategy Learning with N-step Deep Q-learning Network.} In the context of interdependent network dismantling using Deep Reinforcement Learning (DRL), the agent is trained to learn the nodes' removal strategy that maximally disrupts network connectivity across the layers. The Deep Q-Network (DQN) approach addresses this problem by approximating the Q-value function with a deep neural network. The function \( Q(s, a; \theta) \) estimates the expected cumulative reward of taking action \( a \) in state \( s \) with parameters \( \theta \). Here, the action \( a \) represents the selection of a node to remove, and we use node vectors learned from multiplex representation as action embeddings. The state \( s \) represents the current configuration of the multi-layer network. We introduce a virtual node, where all nodes are neighbors of the virtual node, and use the vector of virtual nodes as the state embedding. The reward \( r \) is designed to quantify the impact of node removal on the network's connectivity (see Eq.~\ref{eq:reward}); we measure rewards using the decrease in the size of the LMCC.

Removing a single node has a long-term influence on future states and actions. To capture this long-term effect, we apply the n-step DQN model~\cite{sutton1988learning}. This model aims to minimize the difference between the predicted Q-value and the target Q-value, which includes the cumulative immediate reward and the discounted maximum future Q-value (detailed computation of the Q-value can be found in the Section \ref{decoder}). The agent removes nodes according to the predicted Q-values. As the agent removes nodes and the topology of the multiplex network changes, the embedding vectors of state and action are continuously updated. The node removal process continues until the interdependent network is fully dismantled (see Fig.~\ref{fig:framework} for more details about the training process).

\subsection{Analysis}
We train two MultiDismantler agents with identical training architectures but distinct reward functions to handle node dismantling tasks with 
unit and degree costs.
Also, we consider cost values randomly generated from uniform, normal and Poisson distributions.
We perform tests on synthetic networks generated according to the GMM model.  Also, we analyze 9 real-world multi-layer networks.
We compare our results against various state-of-the-art dismantling methods, including High Degree Adaptive (HDA), Collective Influence (CI), MinSum, FINDER, and NIRM. These algorithms are primarily designed for the dismantling of single-layer networks. Currently, there are no dismantling algorithms specifically tailored for interdependent networks. In this study, we apply these algorithms to each layer individually and then selected the maximum dismantling scores from each layer, sorting them to obtain a dismantling sequence of nodes, see Methods for details.

\subsubsection{Performance on synthetic networks}
\begin{figure}[htbp]
    \centering
    \begin{subfigure}[b]{0.495\textwidth}
        \centering
        \includegraphics[width=\textwidth]{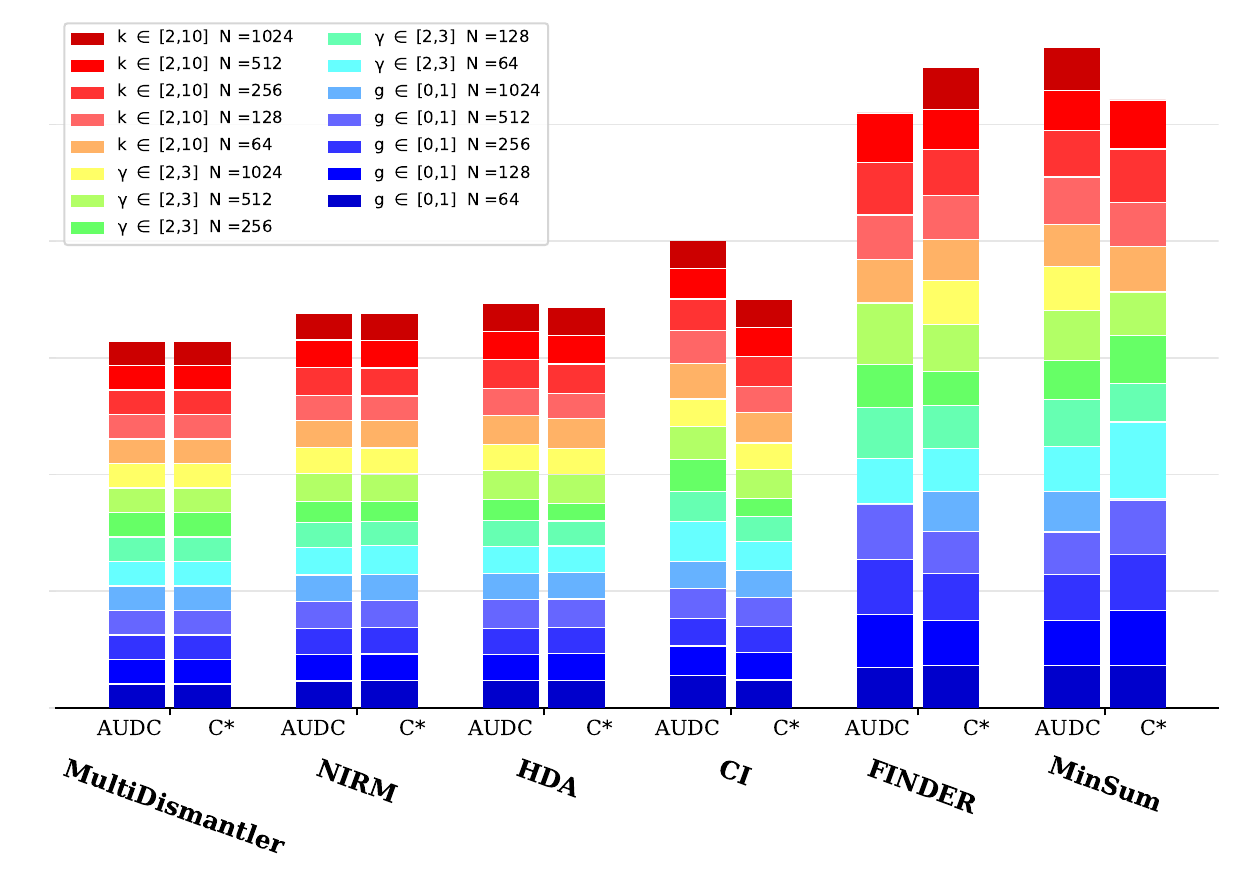}
        \begin{picture}(0,0)
            \put(-126,190){\large A}
        \end{picture}
    \end{subfigure}
    \hspace{0.01cm} 
    \begin{subfigure}[b]{0.495\textwidth}
        \centering
        \includegraphics[width=\textwidth]{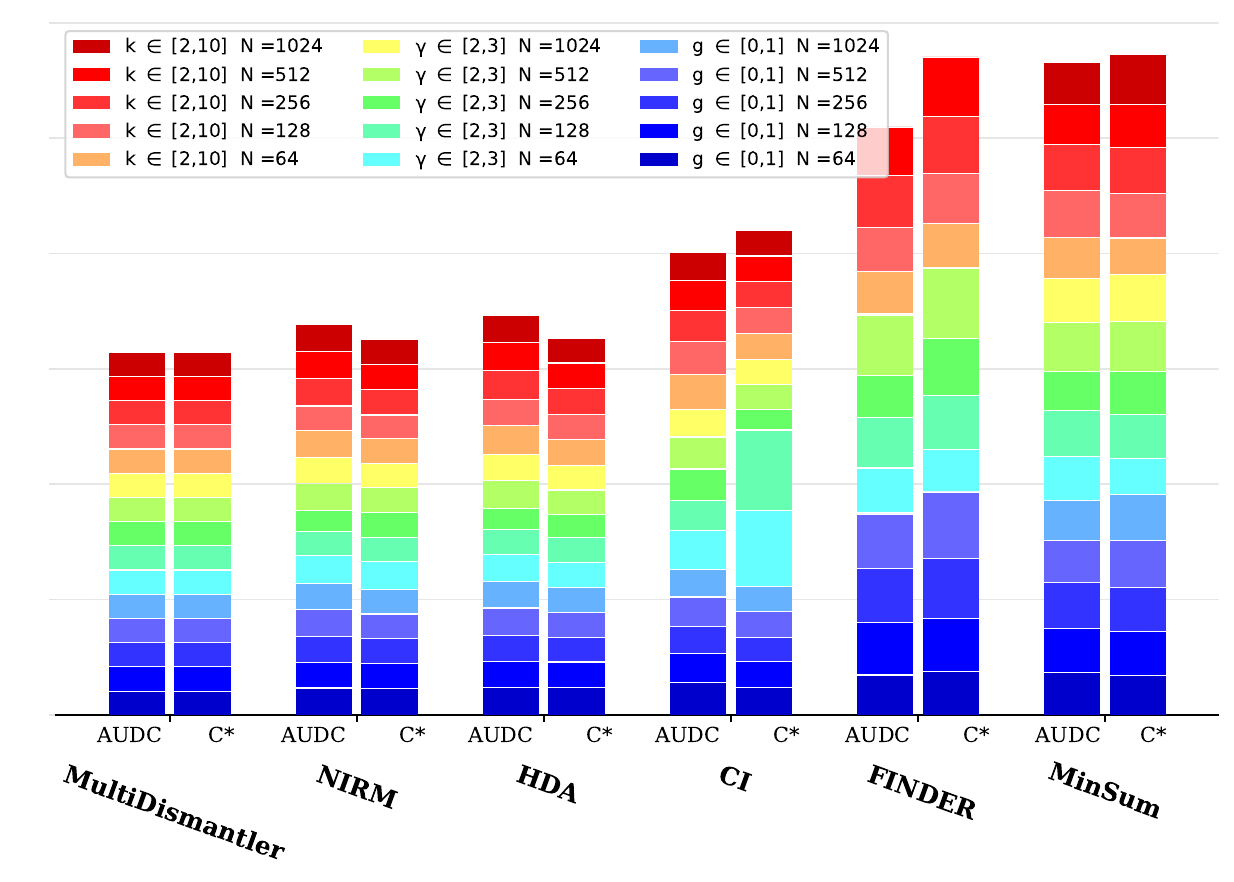}
        \begin{picture}(0,0)
            \put(-126,190){\large B}
        \end{picture}
    \end{subfigure}
    \caption{
    \textbf{Dismantling synthetic interdependent networks.} 
    Dismantling performance under metrics AUDC and $C^*$ for synthetic networks generated using the GMM model with three controllable parameters, i.e., interlayer similarity correlation ($g$), power-law degree distribution exponent ($\gamma$) and expected mean degree ($\overline{k}$). We tune one parameter while keeping the other two parameters static to generate interconnected networks with various properties, see SI Table 3 for the parameter setting of the GMM model. For example, to generate networks under condition $g \in [0,1], N = 1024$, we set $\gamma$ to 2.5, $\overline{k}$ to 6, and randomly generate $g$ values from 0 to 1, each value is the average over 20 different instances to capture the property of interconnected networks with different interlayer similarities. \textbf{A} shows the dismantling performance of unit removal cost under AUDC which quantifies the cumulative area under the dismantling curve and $C^*$ which measures the total cost to disintegrate the network smaller than the square root of the original network size. Each value is scaled to one of the proposed MultiDismantler algorithms for the same network. \textbf{B} shows the dismantling performance of degree removal cost.
Note that some values are clipped (limited) of the FINDER and MinSum algorithms under unit costs to improve visualization, see SI Table 10 for the result in each specific experiment.
    }
    \label{fig:synthetic network performance}
\end{figure}
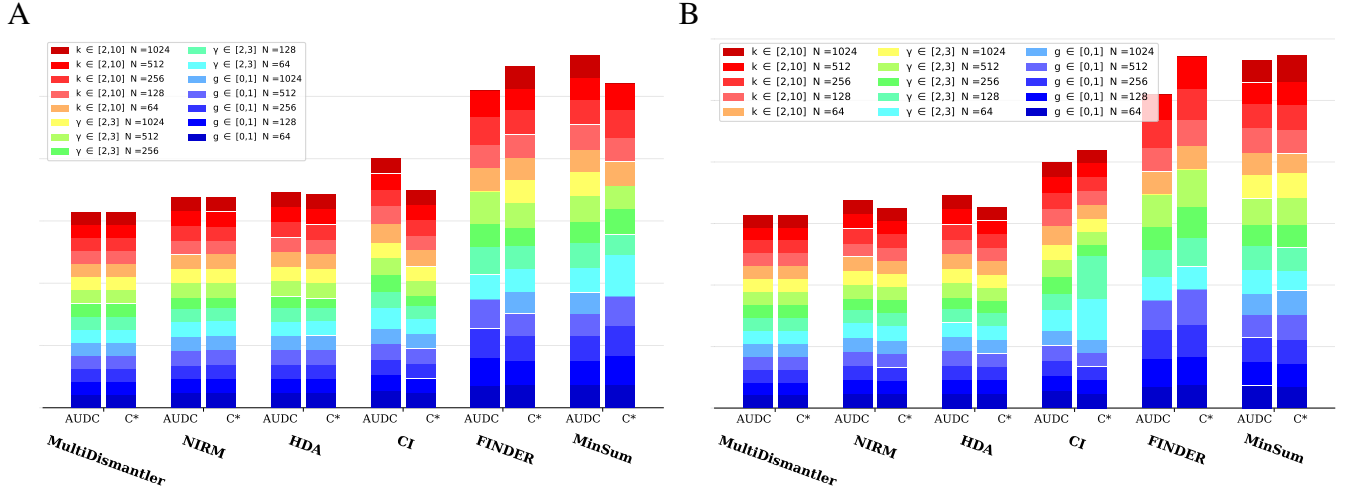

To demonstrate the universal advantages of our algorithm across various types of networks, we vary the main parameters of the GMM model to generate several networks with different intra-layer topologies ($\gamma \in [2,3]$), inter-layer similarities ($g \in [0, 1]$), and network connectivities ($\overline{k} \in [2, 10]$) to model the structure diversity of real-world networks. When we vary one parameter, we keep the other parameters invariant. For each unique combination of network size and GMM parameter settings, we randomly generate 20 networks and reported the results as the average value. Figure \ref{fig:synthetic network performance} shows MultiDismantler's superior dismantling performance on various 
synthetic networks with unit costs and degree costs. SI Tables 10 and 11 highlight that our algorithm outperforms all baseline approaches over all synthetic networks, the AUDC and $C^*$ values are on average 7\% and 10\% lower than the second-best network dismantling algorithm NIRM under unit node removal cost and  8\% and 3.5\% under degree node removal cost. 

\subsubsection{Predicting the ground-truth optimal dismantling sequence}
We evaluate all combinations of nodes to determine the optimal dismantling sequence with minimal AUDC in the unit-cost version of the problem. In this analysis, we consider the network fully dismantled when the LMCC contains only one node.
SI Table 6B employs the Normalized 1-Edit Distance (NED) to quantify the similarity between the dismantling sequences identified by different algorithms and the optimal dismantling sequence computed via brute-force search. SI Table 6A demonstrates that the dismantling sequence computed by MultiDismantler is identical to the optimal sequence and achieves the minimum dismantling AUDC values. 
MultiDismantler shows that machine-learning techniques can well approximate solutions of NP-hard graph combinatorial optimization problems in polynomial time.

\subsection{Results on real-world networks}

\subsubsection{Dismantling performance and inductive learning capacity on real-world networks}
\begin{figure}[htbp]
    \centering
    \begin{subfigure}[b]{0.495\textwidth}
        \centering
        \includegraphics[width=\textwidth]{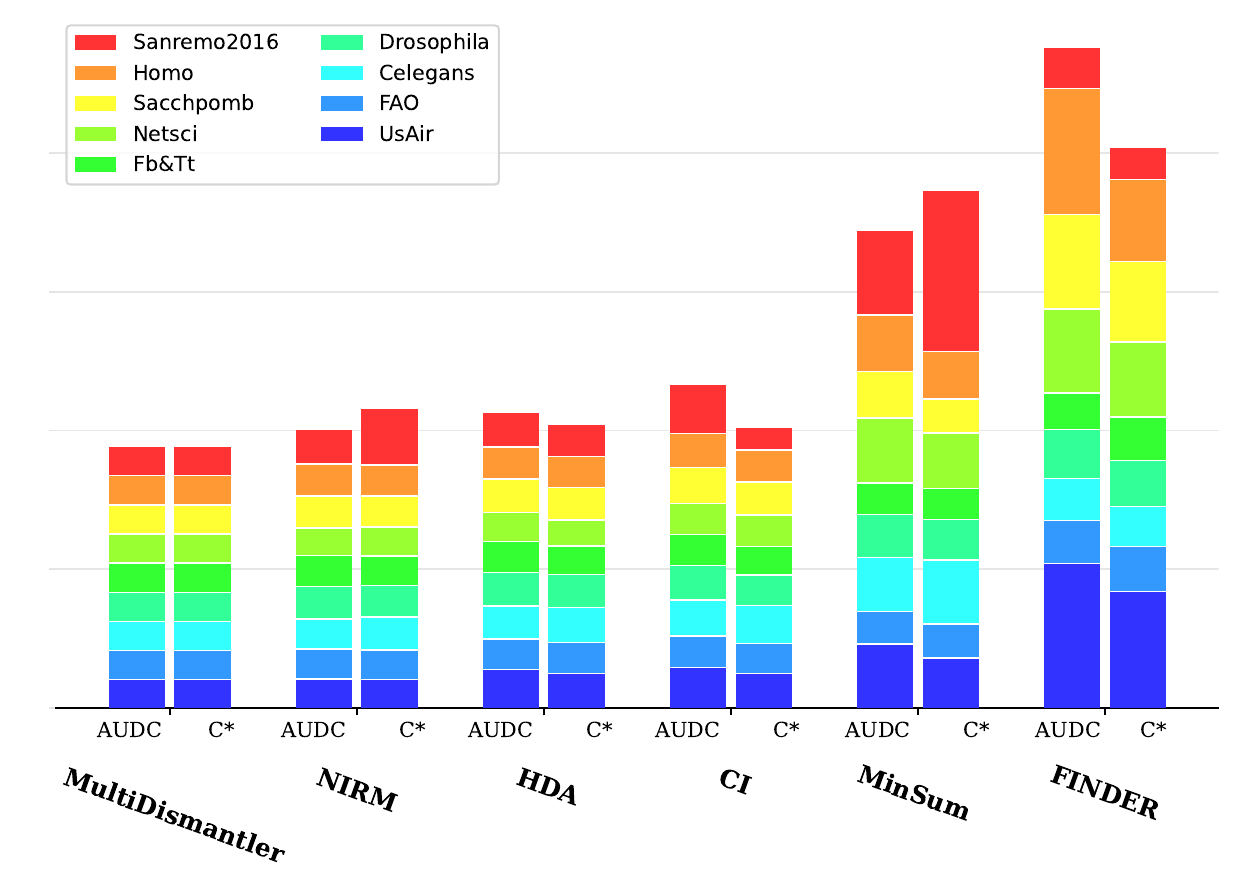}
        \label{fig:image1}
        \begin{picture}(0,0)
             \put(-126,190){\large A}
        \end{picture}
    \end{subfigure}
    \hspace{0.01cm} 
    \begin{subfigure}[b]{0.495\textwidth}
        \centering
        \includegraphics[width=\textwidth]{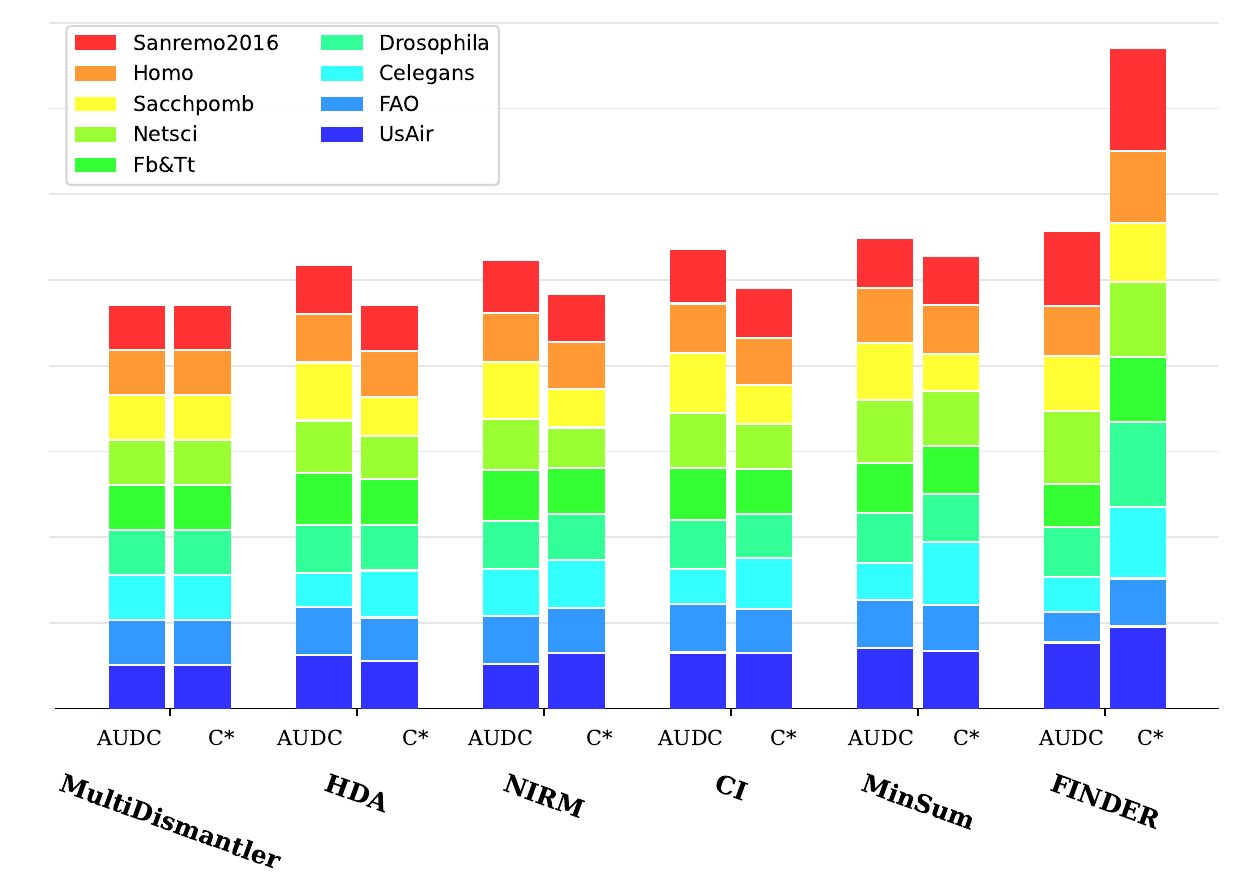}
        \label{fig:image2}
        \begin{picture}(0,0)
           \put(-126,190){\large B}
        \end{picture}
    \end{subfigure}
    \caption{
    \textbf{Dismantling real-world interdependent networks.} 
    Dismantling performance of empirical complex networks from various domains. Each value is scaled to one of the proposed MultiDismantler algorithm for the same network. \textbf{A} shows the dismantling performance comparison under unit removal cost, and \textbf{B} shows the performance for degree removal cost under metrics AUDC and $C^*$. Detailed results for each network are reported in the SI Table 5.
    }
    \label{fig:realworld network performance}
\end{figure}
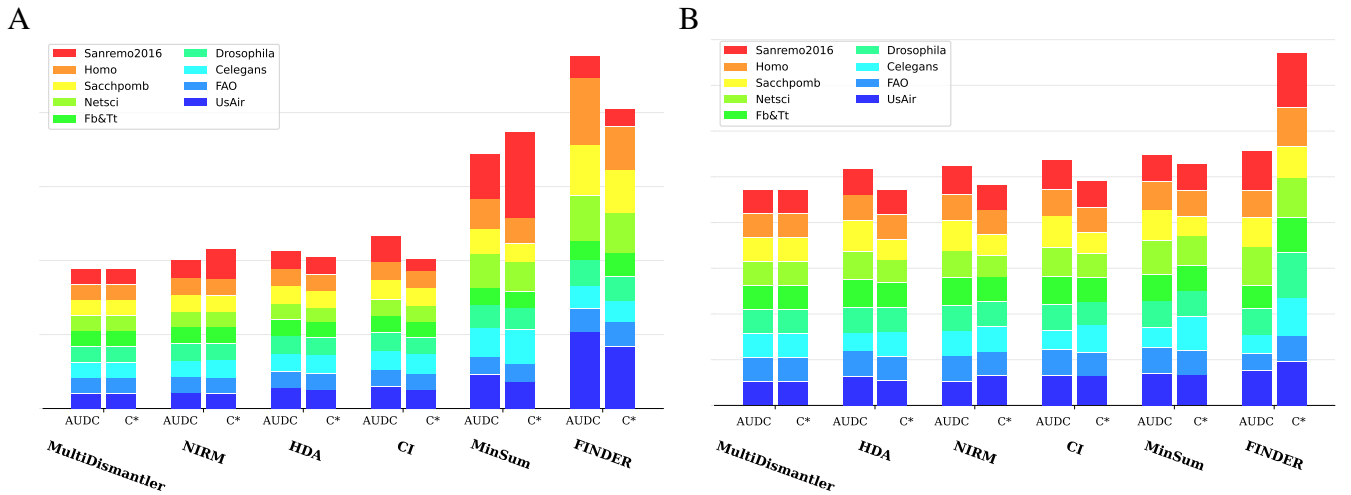

MultiDismantler is trained on a large number of synthetic multiplex networks and then applied real-world networks. We evaluate MultiDismantler on various real-world multiplex networks across different domains (see SI Table 2 for details). Figure \ref{fig:realworld network performance}
shows that MultiDismantler on average outperforms all other methods across multiplex networks with unit and degree removal node costs. For instance, MultiDismantler's cumulative AUDC values are 10\% lower than the second-best NIRM algorithm under degree removal cost and cumulative $C^*$ values are 7\% lower than the second-best dismantling algorithm CI under unit removal cost. We refer the readers to SI Tables 12 for the numerical results of real-world dismantling experiments (more results can be found in SI Table 5).

\begin{figure}[htbp]
\centering
\begin{subfigure}{0.33\linewidth}
  \centering  \includegraphics[width=\linewidth]{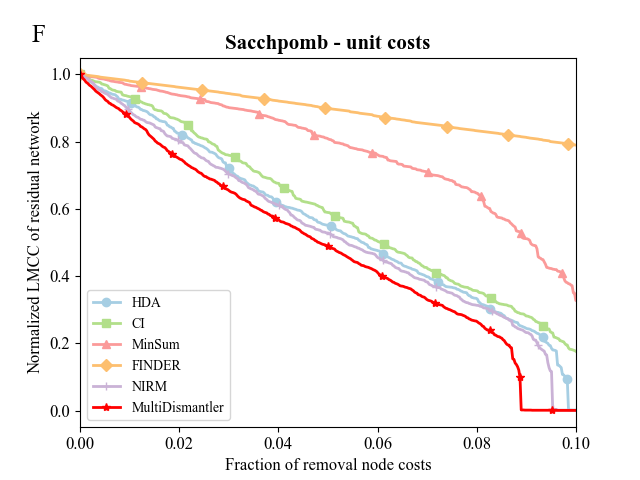}
  \label{Sacchpomb-unit costs }
\end{subfigure}
\hfill
\begin{subfigure}{0.33\linewidth}
  \centering
  \includegraphics[width=\linewidth]{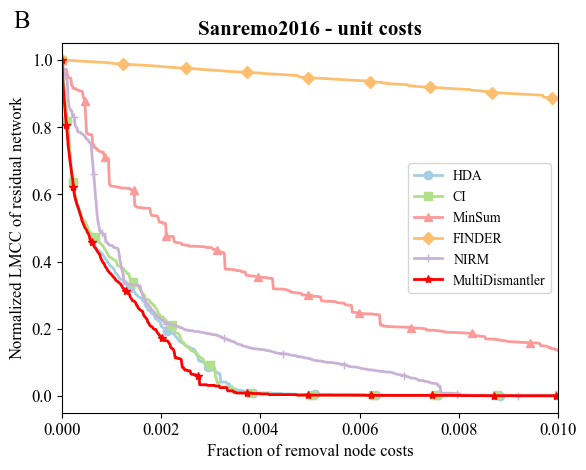}
  \label{Sanremo2016-unit costs}
\end{subfigure}
\hfill
\begin{subfigure}{0.33\linewidth}
  \centering
  \includegraphics[width=\linewidth]{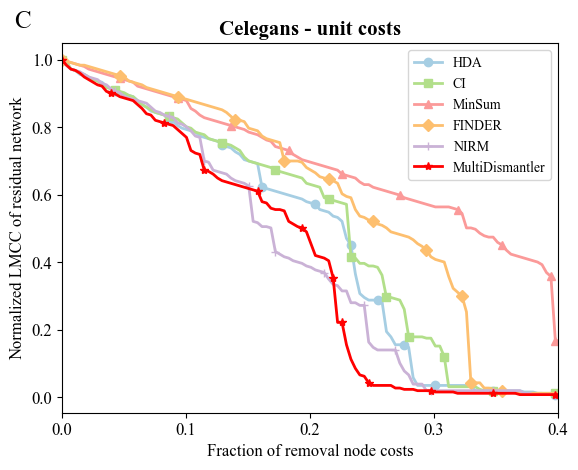}
  \label{Celegans-unit costs}
\end{subfigure}

\begin{subfigure}{0.33\linewidth}
  \centering  \includegraphics[width=\linewidth]{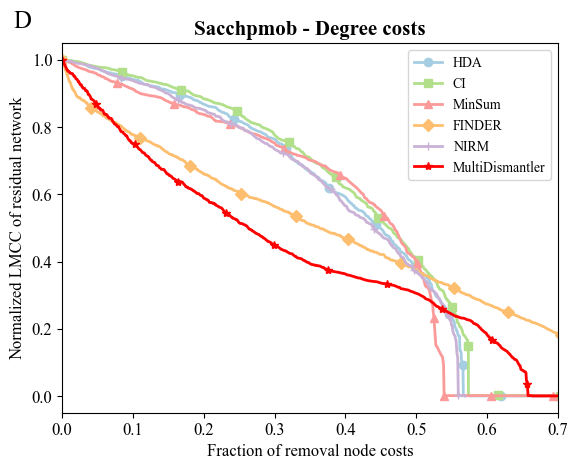}
  \label{Sacchpomb-degree costs}
\end{subfigure}
\hfill
\begin{subfigure}{0.33\linewidth}
  \centering
  \includegraphics[width=\linewidth]{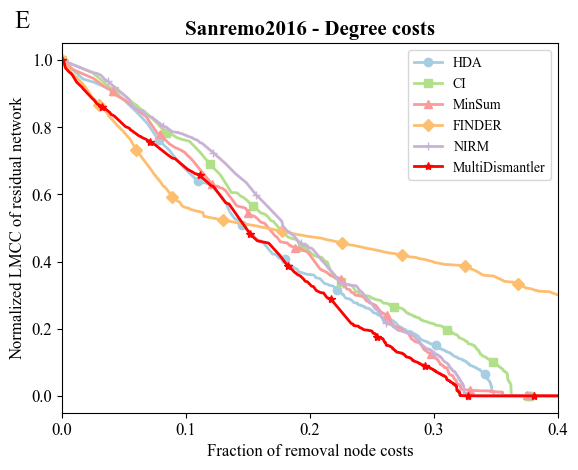}
  \label{Sanremo2016-degree costs}
\end{subfigure}
\hfill
\begin{subfigure}{0.33\linewidth}
  \centering
  \includegraphics[width=\linewidth]{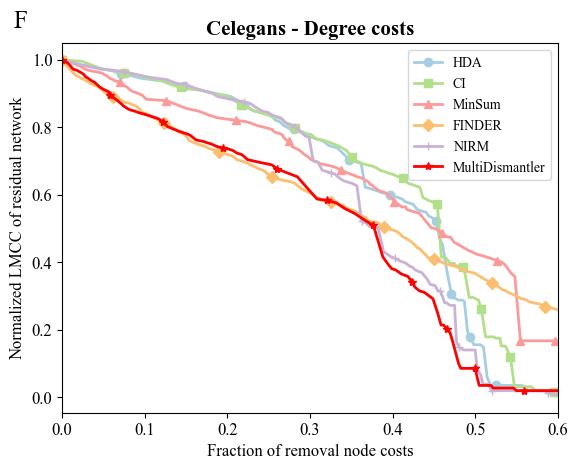}
  \label{Celegans-degree costs}
\end{subfigure}

\caption{The process of dismantling three real multiplex networks by different methods. \textbf{A-C} shows the dismantling of the Sacchpomb network, Sanremo2016 network and Celegans network, respectively, with unit node removal costs. \textbf{D-F} shows the dismantling of the Sacchpomb network, Sanremo2016 network and Celegans network, respectively, with node removal costs based on degree.}
\label{fb-tw-ANC}
\end{figure}

Figure \ref{fb-tw-ANC} illustrates the process of dismantling three real multiplex networks using various methods with unit costs and degree costs. We observe that MultiDismantler consistently achieves the lowest LMCC for the same fraction of removal node costs compared to other dismantling methods on most real multiplex networks, indicating MultiDismanlter is more effective in identifying the dismantling solution (see SI Figure 4 for the dismantling process on real-world networks under unit removal cost). Additionally, in some dismantling curves, our algorithm's curve may initially be higher than those of other methods but then decline more rapidly. This suggests that the model predicts a long-term dismantling strategy: while some target nodes may not directly cause significant network disruption, they are essential in rendering the network vulnerable, making it prone to rapid breakdown upon the removal of other nodes.

To further assess MultiDismantler's effectiveness on networks with weighted removal node costs, we apply the agent trained with random weighted node removal costs. We consider random costs extracted from the uniform, the normal and the Poisson distribution. SI Table 7 shows that MultiDismantler exhibits excellent generalization ability, achieving the best performance on unseen normal and Poisson cost distributions, despite being trained costs randomly extracted from the uniform distribution. This generalization capability of MultiDismantler demonstrates its practical value in real-world scenarios with uncertain node removal costs.

\subsubsection{Analyzing the key components of MultiDismantler}

To better understand the contributions of the various components and determine the essential elements that contribute to the outstanding dismantling performance of MultiDismantler, we systematically alter the three main components under the unit-cost scenario: the attention mechanism in MGNN, the training corpus GMM, and inter-layer angular similarity. SI Table 4A validates the effectiveness of the 
inter-layer attention.
In the first column, we disable the inter-layer message-passing mechanism, resulting in MultiDismantler's inability to fuse representations from different layers, which leads to worse dismantling accuracy. Next, we replace the training corpus from the GMM
with the BA
model with randomly selected interdependent nodes.  The first column in SI Table 4B indicates that MultiDismantler trained with the BA network performs significantly worse than the model trained with GMM. The performance of MultiDismantler degrades even more than for other algorithms, such as HDA
and CI.
Inter-layer similarity also significantly influences MultiDismantler's performance. We adjust the inter-layer 
similarity correlation (\( g \)) of the GMM model and train agents with varying degrees of inter-layer similarity. As shown in SI Table 4C, the performance is rather stable as $g$ is varied and deteriorates only for extreme $g$ values (i.e., $g=0$ and $g=1$).


\subsection{Application to the prevention of disease spreading and the attack-protect task in various real-world networks}

Network connectedness is a crucial factor for network functionality, with the size of LMCC being particularly relevant to problems such as optimal disease spreading\cite{morone2015influence} and attack-protect strategies\cite{albert2000error}. 
Below, we show that MultiDismantler provides a significant advantage in identifying critical nodes to either reduce or maintain network connectivity compared to other dismantling algorithms.

\subsubsection{Evaluating MultiDismantler’s efficacy in immunization strategies}

We leverage MultiDismantler to design immunization strategies for disease spreading on two multiplex social networks: Fb\&Tw network \cite{MagnaniSBP10} and Sanremo2016\cite{2020Unraveling} First, we compute the node dismantling sequence using MultiDismantler with unit cost, identifying nodes that play vital roles in network connectivity. We then determine the minimum number of nodes whose removal reduces the size of the LMCC to 1, and subsequently remove the top half of these nodes. For comparison, we remove the same number of nodes based on their importance rankings as identified by other algorithms. We run Susceptible-Infected-Recovered (SIR) disease-spreading dynamics on the remaining multiplex network, setting all nodes to the susceptible state and randomly selecting one node as seed for spreading. The SIR parameter settings are consistent with those used in the study by Patwardhan et al.\cite{patwardhan2023epidemic}the spreading rate is set slightly higher than the critical one, i.e., $\beta$ = $<k>/<k^2>+0.02$, where $<k>$ and $<k^2>$ are respectively the first and second moment of the average degree of interdependent network. For simplicity, we apply the same spreading rate for all layers. 


Figures \ref{fig:applications-attack} A and B demonstrate that
an immunization strategy based on MultiDismantler
effectively prevents the widespread outbreak of a disease.
Compared to other algorithms, 
the immunization strategy informed by MultiDismantler significantly reduces the number of 
individuals that must be immunized to suppress the spread of an epidemic. 
 
\subsubsection{MultiDismantler's performance in protecting network connectedness under targeted attacks and random failures} 


We protect the first $1\%$ nodes in the sequence of removal identified by MultiDismantler or the other competing methods. We then perturb the network by removing $1\%$ of the remaining nodes either at random (i.e., random-failure protocol) or ranking and removing the first $1\%$ nodes according to the importance defined by MultiDismantler.
We then iterate the above node-protection and node-removal processes until the size of the LMCC reaches 1
or the number of attacked nodes exceeds the number of effective nodes in the residual network (i.e., excluding already deleted and isolated nodes). 
It is worth noting that we select $1\%$ of the initial network's nodes every iteration, rather than $1\%$ of the residual network's nodes.

We report our results in 
Table \ref{table-attack}.
These results demonstrate that MultiDismantler maintains network robustness, outperforming other algorithms under both the targeted-attack and the random-failure scenarios. The adaptability of MultiDismantler to both random and targeted failures underscores its versatility and reliability. By protecting the most critical nodes identified through its dismantling sequence, the algorithm not only mitigates immediate damage but also fortifies the network against subsequent failures. This dual capability of dismantling and protecting networks highlights MultiDismantler's comprehensive approach to network robustness, making it a valuable tool for enhancing the resilience of multiplex networks.

\begin{figure}[htbp]
\centering
\begin{subfigure}[b]{0.49\linewidth}
  \centering  \includegraphics[width=\linewidth]{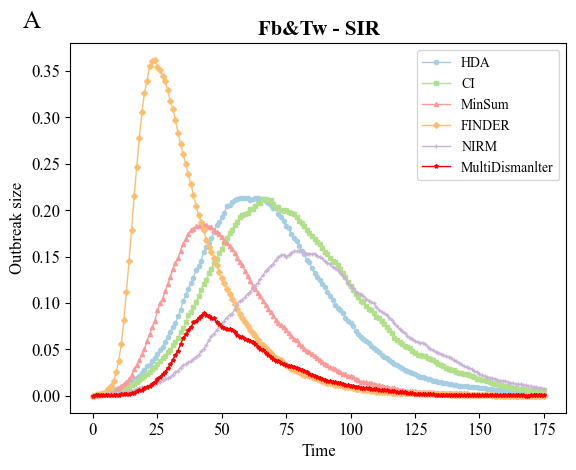}
  \label{fig:Fb&Tw-SIR}
\end{subfigure}
\hfill
\begin{subfigure}[b]{0.49\linewidth}
  \centering
  \includegraphics[width=\linewidth]{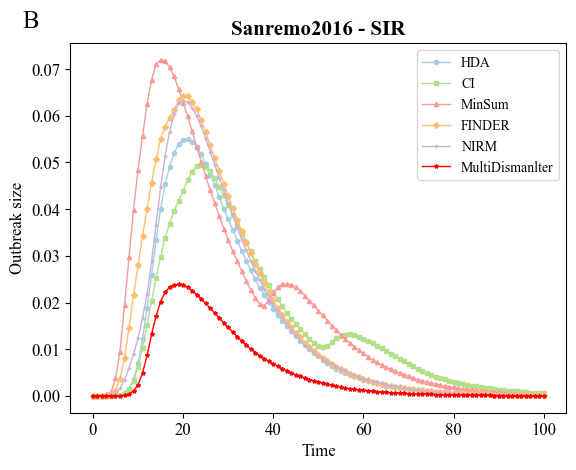}
  \label{fig:Sanremo2016-SIR}
\end{subfigure}
\hspace*{1mm}
\begin{subfigure}[b]{0.49\linewidth}
  \includegraphics[width=\linewidth]{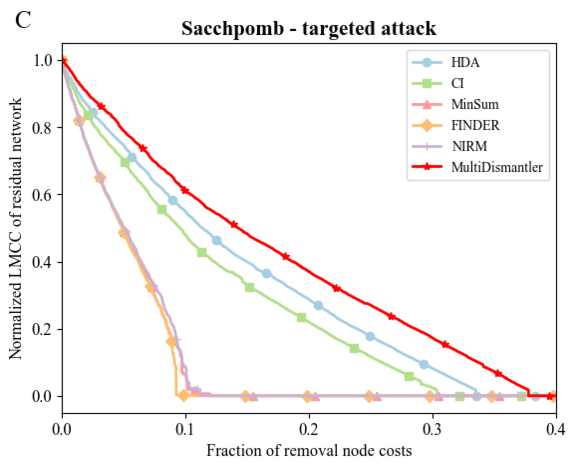}
  \label{fig:targeted_attack}
\end{subfigure}
\hspace*{0.1mm}
\begin{subfigure}[b]{0.49\linewidth}
  \includegraphics[width=\linewidth]{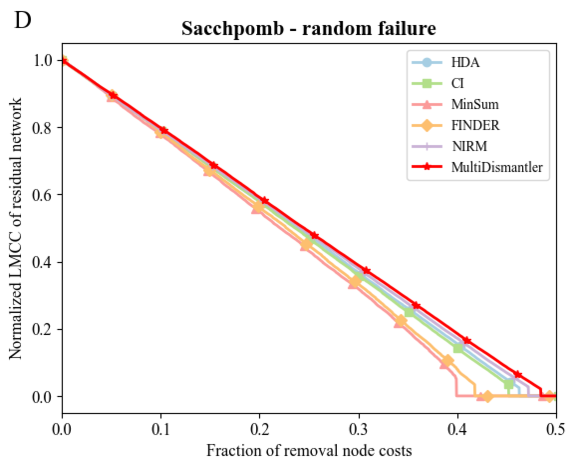}
  \label{fig:random_failure}
\end{subfigure}
\caption{Performance of MultiDismantler in reducing the outbreak size of disease and maintaining network robustness. \textbf{A} and \textbf{B} show MultiDismantler's performance in reducing the outbreak size of disease on offline and online social networks. We simulate SIR dynamics on the remaining social networks after the node removal strategies of different dismantling algorithms to mimic quarantine or immunization of real-world disease control measures. With the same number of removed nodes, the outbreak size under the MultiDismantler controlling strategies is consistently better than other comparison algorithms. \textbf{C} and \textbf{D} simulate network robustness of the Sacchpomb network under targeted attack and random failure. 
Multiplex networks are robust against random failure and MultiDismantler has a slight advantage in maintaining network robustness, }
\label{fig:applications-attack}
\end{figure}

Figures \ref{fig:applications-attack} C and D show the dismantling process of Sacchpomb network with different protection strategies. We observe that MultiDismantler can often achieve the highest  LMCC for the same fraction of removal node costs, whether in the case of targeted attacks or random failures. This suggests that our model is highly effective in optimizing the network's robustness and stability by accurately identifying and protecting the nodes that play crucial roles in maintaining network connectivity.

\begin{table}[htbp]
\caption{Real-word Network robustness under random failure and targeted attack with different protection strategies. For random failure, the results are averaged over 5 trials.  The best results are bold (corresponding the highest AUDC value).}
\centering

\subfloat[Random failure]{
\begin{tabular}{ccccccc}
\hline
Dataset & HDA & CI & MinSum & FINDER & NIRM & MultiDismantler \\ 
\hline
USAir & \textbf{0.420±0.032} & 0.366±0.064 & 0.377±0.046 & 0.343±0.069 & 0.419±0.024 & 0.419±0.024\\\

FAO  & 0.410±0.007 & 0.408±0.006 & 0.406±0.004 & 0.400±0.014 & \textbf{0.413±0.009} & 0.411±0.007 \\

Celegans & 0.379±0.012 & 0.389±0.008 & 0.371±0.014 & 0.386±0.008 & \textbf{0.392±0.010} & 0.390±0.007\\

Drosophila & 0.279±0.011 & 0.277±0.013 & 0.264±0.016 & 0.268±0.007 & 0.272±0.012 & \textbf{0.284±0.011}\\

Fb\&Tt  & 0.463±0.003 & 0.460±0.003 & 0.462±0.003 & 0.461±0.006 & 0.462±0.003 & \textbf{0.465±0.003} \\

NetSci & 0.325±0.005 & 0.318±0.007 & 0.315±0.013 & 0.315±0.017 & 0.317±0.010 & \textbf{0.327±0.009}\\

Sacchpomb & 0.237±0.001 & 0.234±0.003 & 0.218±0.006 & 0.223±0.003 & 0.239±0.004 & \textbf{0.246±0.002} \\

Homo & 0.211±0.001 & 0.207±0.001 & 0.185±0.011 & 0.166±0.010 & 0.203±0.002 & \textbf{0.217±0.001} \\

Sanremo2016 & 0.207±0.001 & 0.202±0.001 & 0.138±0.005 & 0.125±0.011 & 0.198±0.001 & \textbf{0.211±0.001} \\
\hline
\end{tabular}
\label{table-random-attack}
}

\vspace*{0.02\linewidth}

\subfloat[Targeted attack]{
\begin{tabular}{ccccccc}
\hline
Dataset & HDA & CI & MinSum & FINDER & NIRM & MultiDismantler \\ 
\hline
USAir & 0.070 & 0.069 &  0.053 & 0.090 & \textbf{0.136} &\textbf{0.136}\\

FAO  & 0.331 & 0.346 & 0.287 & 0.254 & 0.346 & \textbf{0.350} \\
Celegans & 0.226 & 0.226 & 0.161 & 0.174 & 0.239 & \textbf{0.239}\\

Fb\&Tt  & 0.326 & 0.326 & 0.332 & 0.326 & 0.338 & \textbf{0.348} \\

Drosophila & 0.117 & 0.112 & 0.057 & 0.060 & 0.117 & \textbf{0.132} \\

Sacchpomb & 0.135 & 0.116 & 0.051 & 0.049 & 0.052 & \textbf{0.162} \\

NetSci & \textbf{0.132} & 0.111  & 0.063 & 0.057 & 0.130 & 0.122\\

Homo & 0.127 & 0.109  & 0.059 & 0.029 & 0.097 & \textbf{0.144} \\

Sanremo2016 & 0.172 & 0.156 & 0.006 & 0.002 & 0.174 & \textbf{0.182} \\
\hline
\end{tabular}
\label{table-target-attack}
}
\label{table-attack}
\end{table}

\section{Conclusions}

In this study, we address the problem of dismantling interdependent multi-layer networks using a deep-learning-aided approach. Our proposed approach, MultiDismantler, incorporates a multiplex embedding framework to encode networks and a DQN-based decoder to effectively train agents in learning complex dismantling strategies. Extensive experiments on both synthetic and real-world networks demonstrate the superior performance of MultiDismantler compared to existing state-of-the-art dismantling algorithms. We stress that none of the existing algorihtms is designed to dismantle multi-layer networks, rather they represent adaptations of algorithms designed for single-layer networks. On small networks, the dismantling sequence produced by MultiDismantler matches the optimal sequences computed using brute-force search, highlighting the potential of MultiDismantler in well approximating solutions of an NP-hard problem in polynomial time.

MultiDismantler exhibits remarkable generalization capabilities across various network topologies and node removal cost distributions, meaning that it can be safely applied to previously unseen networks. Further, although MultiDismantler serves to approximate solutions to the network dismantling problem, it turns out useful also in designing solutions of related problems, such as the prevention of disease spreading in social networks and the attack-and-protect task in interdependent critical infrastructures.

Our work underscores the potential of combining graph neural networks with reinforcement learning for complex network optimization tasks and opens avenues for future research aimed at enhancing the scalability and adaptability of such models for broader applications.

\section{Methods}


\subsection{Dismantling multi-layer interdependent networks}

\subsubsection{Problem formulation}
\label{Problem formulation}
We consider a network 
consisting of  two layers of interactions. Nodes in the two layers  are one-to-one
interdependent. For simplicity of description but without loss of generality, we do not make a formal distinction between one node and its replica in the other layer, rather we act as the two layers would share the same set of nodes. Thus, we denote the network as \( G = (V, E^{(1)}, E^{(2)}) \), where \(V\) is the set of nodes shared across the layers, with $|V| = N$ denoting the total number of nodes, and \(E^{(\ell)}\) are the  edges in layer $\ell=1, 2$.
The generic element of the adjacency matrix of layer $\ell$ is $A_{ij}^{(\ell)} =1$ if nodes $i$ and $j$ are connected, or $A_{ij}^{(\ell)} =0$ if they are not connected. 

Large-scale connectedness of the network is quantified in terms of the fraction of nodes that belong to the largest mutually connected component  (LMCC) of the network as $P_{\infty} = \frac{N_{LMCC}}{N_I}$, where $N_{LMCC}$ is the number of nodes in the LMCC and $N_I$ is the number of nodes in the initial LMCC, when all $N$ nodes are present in the network~\cite{buldyrev2010catastrophic}.
The size of the LMCC can be reduced by removing nodes belonging to a subset $S \subseteq V$.  Based on our defintions, if $S = \emptyset$, then $P_{\infty} (\emptyset) = 1$; if $S \neq \emptyset$, then $P_{\infty}(S) \leq 1$; also, $P_{\infty}(V) = 0$.

Network dismantling, often referred as the optimal percolation problem, can be seen as the constrained minimization problem
\begin{equation}
    S^* (C) = \arg \, \min_{S | F(S) = C} P_{\infty}(S) \; .
    \label{eq:opt_perc}
\end{equation}
The constraint is imposed on the value of the cost function 
$F(S)$ of removing elements of the set $S$. 
In this paper, we consider three main types of cost functions:
(i) unit cost; (ii) degree cost; (iii) random cost.
In the unit-cost version of the problem, the cost function associated to the set $S$ equals its size, i.e., $F(S) = |S|$. The degree-cost function of variant (ii) is defined as $F(S) = \sum_{s \in S} \, \sum_{\ell =1}^2 \, k_s^{(\ell)} - \sum_{s,t \in S} A_{st}^{(\ell)}$, where $k_s^{(\ell)} = \sum_{j \in V} A_{sj}^{(\ell)}$  is the degree of node $s$ in layer $\ell$, the sums run over all nodes in the set  $S$, and edges shared by nodes within the set $S$ are counted only once. In the random-cost variant, the cost associated with each node $s$
is $F(S) = \sum_{s \in S} \, \sum_{\ell =1}^2 u_s^{(\ell)}$ where $u_s^{(\ell)}$
is extracted at random from the uniform distribution defined in the interval $[0,1]$, the normal distribution with average $0.5$ and variance $0.1$, or the Poisson distribution with average equal to $5$.

An important aspect in the characterization of the optimization problem is the identification of the minimum-cost set of nodes able to lead to the disappearance of a macroscopic LMCC. Such a condition is defined in the problem
\begin{equation}
    S_c = \arg \min_{S | P_\infty(S) \leq 1/\sqrt{N_I}} F(S) \; .
    \label{eq:opt_perc2}
\end{equation}
Essentially, only sets $S$ that can reduce $P_\infty$ below the conventional threshold value $1/\sqrt{N_I}$ are considered as potential solutions to the problem~\cite{clusella2016immunization}.

\subsubsection{Approximating solutions of the network dismantling problem}

 All algorithms considered in this paper construct approximate solutions to the dismantling problem sequentially, meaning that the set corresponding to the proposed solution is built by adding one element at a time. Indicate with $r_1, r_2, \ldots, r_e, \ldots, r_N$  the labels of the ranked nodes according to the algorithm at hand. Then, define $\tilde{S}_t = \bigcup_{e =1}^t \{r_e\}$, i.e., the approximate solution of the algorithm when the set is composed of exactly $t$ elements. By definition, $\tilde{S}_0 = \emptyset$ and $\tilde{S}_N = V$.  We clearly have that $P_\infty ( \tilde{S}_{t-1} ) \geq P_\infty ( \tilde{S}_t )$ and $F(\tilde{S}_t) \geq F(\tilde{S}_{t-1})$ for all $t=1, \ldots ,  N$.

We evaluate the performance of an approximate algorithm by measuring the area under the dismantling curve (AUDC) as
\begin{equation}
\text{AUDC} = 
\frac{1}{F(V)} 
\, \sum_{t = 1}^{N} P_\infty ( \tilde{S}_t ) \, \left[ F(\tilde{S}_t) - F(\tilde{S}_{t-1}) \right]
\; ,
    \label{eq:evaluation}
\end{equation}
where $F(V)$ is the cost associated to the removal of all nodes from the graph.
This is the generalization of the the so-called robustness metric introduced by Schneider {\it et al.}~\cite{schneider2011mitigation}. For computational reasons, we approximate AUDC by summing only the first $N_c$ contributions such that $P_\infty ( \tilde{S}_t ) = 1/N_I$  for $t = 0, . . . , N_c$. This represents a good approximation of Eq.~\ref{eq:evaluation}.

Also, we evaluate the performance of an approximate algorithm to solve the problem of Eq.~(\ref{eq:opt_perc2}) by measuring $F(\tilde{S}_c)$, where $\tilde{S}_c = \arg \min_{\tilde{S}_t | P_\infty(\tilde{S}_t) \leq 1/\sqrt{N_I}} F(\tilde{S}_t)$. To make the metric comparable across networks and/or variants of the dismantling problem, we define the dismantling cost as 
\begin{equation}
C^* = \frac{F(\tilde{S}_c)}{F(V)}
\; .
 \label{eq:evaluation_a}
\end{equation}

Note that both AUDC and $C^*$ are defined in the interval $[0,1]$. Also for both metrics, low values indicate good performance of the solution of the dismantling problem, whereas values close to one denote poor performance.


\subsection{MultiDismantler}
We provide below details on enconder, decoder, loss function and dismantling strategy used in MultiDismantler. 

\subsubsection{Encoder}
The framework of the Multiplex Graph Neural Network is shown in Supplementary Figure 5. MGNN consists of an intra-layer feature convolutional operation and an inter-layer representation integration process. 

Node feature initialization in network embedding is vital for providing a beneficial starting point and accelerating convergence speed, inspired by the feature initialization operation of FINDER~\cite{fan2020finding}, MGNN initialized node feature by transforming the input feature through a one-layer multilayer perceptron

\begin{equation}
\label{eq:init}
\textbf{h}_v^{0,(\ell)}\,=\,Norm(\text{ReLU}(\textbf{W}_1 \cdot \textbf{X}_v^{(\ell)})) \; , 
\end{equation}

where $\textbf{X}_v^{(\ell)}$ is the normalized degree of node $v$ in layer $\ell$ (dividing the degree of a node by the maximum degree in the layer), \text{ReLU}(·) = max(0, ·) is the nonlinear activation function, $\textbf{W}_1$ is a trainable weight matrix, and $Norm($·$)$ denotes the L2 normalization.

After obtaining the initialized features, we employ an inductive transformation function for each node in each layer $\ell$ to iteratively aggregate the node’s prior representation with the embeddings of its neighbors. The neighbor message-passing process is

\begin{equation} 
\label{eq:intra_aggregate}
\textbf{h}_v^{k, (\ell)}\,\leftarrow\,Norm(\text{ReLU}(\textbf{W}_4\cdot(\{\textbf{W}_3\cdot\textbf{h}_v^{k-1, (\ell)} \}\cup\{\textbf{W}_2\cdot\sum\nolimits_{j \in N(v)^{(\ell)}}\textbf{h}_j^{k-1, (\ell)}\}))) \; .
\end{equation}

 In Eq.~\ref{eq:intra_aggregate},  $\textbf{h}_v^{k, (\ell)}$ represents the learned embedding of node $v$ in layer $\ell$ for the $k-$th forward propagation step, which captures $k$ hops neighbors’ information, $\textbf{W}_2$, $\textbf{W}_3$ and $\textbf{W}_4$ are trainable weight matrices, $\cup$ denotes the concatenation operation, and $N(v)^{(\ell)}$ denotes the set of neighbors of node $v$ in layer $\ell$. The final intra-layer node representation can be represented as ${h}_v^{(\ell)}$. 

In multiplex networks, nodes from different layers interact and influence each other. To quantify cross-layer influence and address the heterogeneity of intra-layer node representations, we map node embeddings from different layers to a unified embedding space using 

\begin{equation}
\label{eq:mapping}
\textbf{h}_v^{(\ell)}\,=\,\tanh(\textbf{W}_5 \cdot \textbf{h}_v^{(\ell)} + \textbf{b}_1) \; ,
\end{equation}

where $\textbf{W}_5$ is a trainable weight matrix, $\textbf{b}_1$ denotes a trainable bias vector, and $\tanh$ is the nonlinear activation function.


Previous aggregation algorithms typically employ fixed cross-layer weights to quantify inter-layer influence, overlooking the variability of inter-layer connection weights even within the same layer. To capture these high-order cross-layer node interaction relations and enhance MGNN's encoding expressiveness, we designed a node-level attention mechanism that computes custom attention weights for every cross-layer relation, i.e.,

\begin{equation}
\label{eq:bitwise} 
\alpha_v^{(\ell\leftarrow q)}\,=\, \frac{e^{\text{Sigmoid}\left({\textbf{W}_6} \cdot\left( \textbf{h}_v^{(\ell)}\otimes \textbf{h}_v^{(q)}\right)+ \textbf{b}_2\right)}} {\sum_{p=1}^{P}e^{ \text{Sigmoid}\left(\textbf{W}_6\cdot\left( \textbf{h}_v^{(\ell)}\otimes\textbf{h}_v^{(p)}\right) + \textbf{b}_2\right)}}
\end{equation}
where $\alpha_v^{(\ell\leftarrow q)}$ is the computed inter-layer attention of node $v$ from layer $q$ to $\ell$, $\textbf{W}_6$ is a trainable weight matrix, $\textbf{b}_2$ denotes a bias vector, $P$ is layer number of multiplex network, and $\otimes$ stands for Hadamard product. It is worth noting that $a_v$ is asymmetric, i.e., $a_v^{(\ell\leftarrow q)}$ is unnecessarily equal to $a_v^{(q\leftarrow \ell)}$, which indicates that node in one layer can be important to another layer, but not vice versa.

MGNN aggregates the learned intralayer representation and inter-layer weights influence to form the final node embedding using 

\begin{equation}
\label{eq:inter_aggregate}
\textbf{z}_v^{(\ell)}\,=\,\textbf{h}_v^{(\ell)}\,+\,\sum\nolimits_{p=1,p \neq \ell}^{P} \alpha_v^{(\ell\leftarrow p)} \cdot \textbf{h}_v^{(p)} \; ,
\end{equation}
where $\textbf{z}_v^{(\ell)}$ is the final representation of node $v$ in layer $\ell$.

\subsubsection{Decoder}
\label{decoder}
The decoder framework takes the network state and node action vector as input and computes Q-values (the expected returns) for all possible actions with a neural network. In this paper, we apply node representations learned from MGNN as action vectors, $\textbf{z}^{(\ell)}_{v}$ is the action vector of node $v$ in layer $\ell$. Network state represents the current network topology after the removal of nodes, considering networks are distributed systems, each node only has a partial view of the network, so we add a virtual node $s$ to denote the state in each layer.  In layer $\ell$, all residual nodes are directed neighbors of $s$ but node $s$ is not a neighbor to any other node in that layer. This design helps to avoid over-smoothing and provides a comprehensive view of the network state. We use $\textbf{z}_s^{(\ell)}$ to denote the embedding vector of the virtual node  $s$ in layer $\ell$. We then utilize a two-layer MLP to model the expected cumulative layer-level reward (Q-value) of an action under a given network state with 
\begin{equation}
\label{eq:get single layer Q}
\textbf{Q}^{(\ell)}(s, a_v)\,=\,\textbf{M}_{1} \cdot \text{ReLU}( \textbf{z}^{(\ell)}_{s} \times \textbf{z}^{(\ell)}_{v} \cdot \textbf{M}_2)   \; ,
\end{equation}
where $\textbf{Q}^{(\ell)}(s, a_v)$ denotes the Q-value action $a_v$, i.e., removing the node $v$, under layer state $s$, $\textbf{M}_1$ and $\textbf{M}_2$ are trainable weight matrices, and $\times$ denotes the outer product operation which can model fine dependencies between states and actions\cite{fan2020finding}.

Nodes in different layers have various expected rewards and $\textbf{Q}^{(\ell)}(s, a_v)$ reflects only the partial expected reward of removing node $v$, for example, node $v$ might serve as a core node connecting different communities in one layer but act as a peripheral node in another layer. To have a comprehensive view of node importance from various layers, we employ a layer-level attention mechanism to quantify the final node removal return with 
\begin{equation}
\label{eq:get weight} 
\omega^{(\ell)}\,=\, \frac{e^{\textbf{M}_4\cdot\text{ReLU}\left({\textbf{M}_3} \cdot\textbf{z}_s^{(\ell)}\right)}} {\sum_{p=1}^{P}e^{\textbf{M}_4\cdot\text{ReLU}\left({\textbf{M}_3} \cdot\textbf{z}_s^{(p)}\right)}} \; ,
\end{equation} 
 where $\omega^{(\ell)}$ denotes the reward importance of layer $\ell$, and $\textbf{M}_3$ and $\textbf{M}_4$ are trainable matrices.

We aggregate the final expected return of node $v$ across all layers in 

\begin{equation}
\label{eq:final Q}
\textbf{Q}(s, a_v)\,=\,\sum\nolimits_{p=1}^{P} \omega^{(p)} \cdot \textbf{Q}^{(p)}(s, a_v) \; ,
\end{equation}

where $\textbf{Q}(s, a_v)$ denotes the Q value for removing node $v$ under the state of the entire network.

\subsubsection{Loss Function and Dismantling Process}

For the observed reward we use the impact of node removal on the real network’s connectivity to quantify the direct reward of removal costs as 

\begin{equation}
\label{eq:reward}
r(v) = P_\infty( \{v\}) \, F(\{v\}) \;, 
\end{equation}
where $P_\infty( \{v\})$ is the relative size of the LMCC of the graph when node $v$ is removed, and 
$F(\{v\})$ is the cost of removing node $v$.

The impact of removing a node might not be immediately apparent and can have delayed effects on the network's connectivity. We apply n-step DQN to put n-step $(s_t, a_t, r_t, \ldots, r_{t+n-1}, s_{t+n})$ in a replay buffer to better capture the delayed effects by considering the cumulative target reward over multiple steps. The target Q-value is composed of the cumulative observed rewards, $r_{t, t+n}$ 
and the estimated maximum future reward from the next $t+n$ state under potential action $a^{\prime}$ denotes as $\max _{a^{\prime}} \hat{\textbf{Q}}(s_{t+n}, a^{\prime})$. The decoder framework evaluates an agent's predicted rewards of action $a$ under state $s$ in time $t$, denoted as $\textbf{Q}(s_{t}, a_{t})$. The Q-learning loss aims to minimize the difference between the predicted Q-value and the target Q-value
\begin{equation} 
\label{eq:loss Q func}
Loss_{Q}=\mathbb{E}_{\left(s_{t}, a_{t}, r_{t, t+n}, s_{t+n}\right) \sim U(B)}\left[\left(r_{t, t+n}+\rho \max _{a^{\prime}} \hat{\textbf{Q}}\left(s_{t+n}, a^{\prime} ; \hat{\Theta}_{N}\right)-\textbf{Q}\left(s_{t}, a_{t} ; \Theta_{N}\right)\right)^{2}\right]\\
\; ,
\end{equation}
where $U(B)$ denotes the uniform sampling distribution over the experience replay buffer $B$,  ${\Theta}_{N}$ is current network parameters and $\hat{\Theta}_{N}$ is the target network parameters which are updated periodically from ${\Theta}_{N}$, and $\rho$ is the discount factor that determines the importance of future rewards.

We also add a graph reconstruction loss\cite{wang2016structural} to preserve the network topology information with 
\begin{equation}
\label{eq:loss R func}
Loss_{R}=\sum\nolimits_{p=1}^{2}\sum\nolimits_{i, j=1}^{N} A^{(\ell)}_{ij}\left\|\textbf{z}_i^{(\ell)}-\textbf{z}_j^{(\ell)}\right\|_{2}^{2} \\ \; ,
\end{equation}
$\|\cdot\|_{2}^{2}$ is the squared L2 norm.

The total loss $Loss_T$ is composed of the Q-learning loss $Loss_Q$ and the graph reconstruction loss $Loss_R$ is 
\begin{equation}
\label{eq:loss T func}
Loss_T = Loss_{Q} + \beta Loss_{R} \;,
\end{equation}
where $\beta$ is a hyperparameter that balances the importance of two losses.

Building on the previously established encoding, decoding, and optimization framework, we employ a greedy selection procedure to determine the optimal action and node to remove based on the Q-value. During the training phase, an $\epsilon$-greedy strategy is used, where the action with the highest Q-value is selected with a probability of $(1 - \epsilon)$, while a random action is chosen with probability $\epsilon$. The value of $\epsilon$ is gradually reduced from 1 to 0.05 over 10,000 episodes to balance exploration and exploitation. In the application phase, the well-trained agent model is used to compute Q-values for each node. The Q-value sequence for state and action is represented as \((Q(s,a_1), Q(s,a_2), \ldots, Q(s,a_m))\), where $m$ is the number of remaining nodes. In the application phase, we always select the actions with the highest Q-values to remove. The node selection and removal process continues until the removal results in the disappearance of a macroscopic LMCC.

\subsection{Other dismantling algorithms}
As we mentioned, there are no other algorithms that are specifically designed to dismantle multi-layer interdependent networks. We therefore generalized dismantling algorithms from single- to multi-layer networks.

\subsubsection{High degree adaptive (HDA)}
We assign to each node $i$ the score $s_i = \max(k_i^{(1)},k_i^{(2)})$, and $k_i$ is the degree of node $i$, which ranks nodes solely by degree. We remove node with the highest degree. In cases where nodes share identical degrees, we randomly select and remove one of these nodes. Further, once a node is removed from the network, the degrees of the remaining nodes are updated~\cite{chen2009efficient}.

\subsubsection{Collective Influence (CI)}
We use the adaptive version of the so-called collective influence (CI) centrality~\cite{morone2015influence}. In each layer, the score assigned to each node $i$ is a function of the number and degree of other nodes at distance $v$ from $i$. $v$ is a tunable parameter. For $v=0$, the metric reduces to HDA. Here, we use $v=1$. In our experiments, we assign to a given node a score equal to the maximum of its CI scores in the two layers. Further, once a node is removed from the network, the scores of the remaining nodes are updated. The implementation of the CI algorithm we applied can be downloaded \href{https://github.com/zhfkt/ComplexCi}{here}.

\subsubsection{MinSum}
We first apply the MinSum~\cite{braunstein2016network} algorithm to each layer of the interconnected network to obtain their dismantling sequences. The position of node $i$ in the dismantling sequence of layer $\ell$ is represented as $r_i^{(\ell)}$. We then assign each node $i$ a score $s_i=\max(1/r_i^{(1)},1/r_i^{(2)})$ and remove node with the biggest score. For nodes with the same score, we randomly select and remove one of those nodes. The implementation of the MinSum algorithm we applied can be downloaded \href{https://github.com/abraunst/decycler}{here}.

\subsubsection{FINDER}
FINDER uses a deep reinforcement learning framework to identify critical nodes for single-layer networks~\cite{fan2020finding}. A well-trained FINDER model estimates q score for each node $i$ in a layer, $q_i^{(\ell)}$ represents the expected return if remove node $i$ in layer $\ell$ for each node in one layer. To apply FINDER in multi-layer interdependent networks, we assign to each node $i$ the score $s_i=\max(q_i^{(1)},q_i^{(2)})$ and remove node with the biggest scores. Notably, we utilized the well trained FINDER version in the unit costs scenario, the FINDER implementation can be downloaded \href{https://github.com/faraz2023/FINDER-pytorch}{here}. As for other weighted scenarios such as degree and random removal costs, we retrained the FINDER model with the same training iterations with Multidismantler for a fair comparison.

\subsubsection{NIRM}
The Neural Influence Ranking Model (NIRM) uses both local and global scoring mechanisms to learn the final scores for each node in a network layer. The final score of node $i$ is given by $s_i=\max(m_i^{(1)},m_i^{(2)})$, where $m_i^{(\ell)}$ denotes the score of node $i$ in layer $\ell$. For nodes with identical final scores, we randomly select and remove one of those nodes. The implementation of the NIRM algorithm we applied can be downloaded \href{https://github.com/JiazhengZhang/NIRM}{here}.

\bibliography{sample}

\section*{Acknowledgements}
This work was supported by grants from the Beijing University of Chemical Technology (grant number 11170044127 and ZY2412); partial support was also received from the Air Force Office of Scientific Research (grant numbers FA9550-21-1-0446 and FA9550-24-1-0039). The funders had no role in study design, data collection, and analysis, the decision to publish, or any opinions, findings, conclusions, or recommendations expressed in the manuscript.

\section*{Author contributions statement}
W.W, and F.R. wrote the paper. W.W, C.Y and F.R. designed the model and experiments. W.W. C.Y and L.L performed the experiments. C.Y, L.L and J.H plotted the figures.

\section*{Code availability statement}
The code developed for this research is available at \url{https://codeocean.com/capsule/6460456}.

\section*{Data availability statement}
All data used in this paper is publicly available. To facilitate the reproducibility of our results, we shared all necessary data at \url{https://codeocean.com/capsule/6460456}.

\end{document}


\renewcommand{\thesubtable}{\Alph{subtable}}

\renewcommand{\figurename}{Supplementary Figure}
\renewcommand{\tablename}{Supplementary Table}

\renewcommand{\theequation}{S \arabic{equation}}

\newcommand{\newchange}[1]{{\color{red}#1}}
\newcommand{\change}[1]{#1}

%


\tableofcontents   
\newpage              

\section{Comparative Analysis: FINDER and MultiDismantler}
\label{compare}
MultiDismantler builds on foundational ideas from FINDER and employs a deep Q-learning model for network dismantling. At the same time, MultiDismantler introduces several critical innovations specifically designed for the dismantling of interdependent networks. In the following, we outline the key differences and similarities between the two methods. Also, we empirically show that the performance's gap between the two methods is rooted in their technical differences.
\subsection{Differences}
\textbf{Generative model for training data}.
FINDER uses synthetic single-layer networks, such as those generated by the Barabási-Albert (BA) model, for training. These networks lack the ability to represent interdependencies and/or geometric correlations across multiple layers, limiting their applicability to multi-layer network problems. In contrast, MultiDismantler employs the Geometric Multiplex Model (GMM), which generates multi-layer networks with realistic cross-layer correlations. This allows MultiDismantler to train on networks that accurately reflect the properties of real interdependent systems.

\textbf{Multiplex Graph Neural Network (MGNN)}.
FINDER’s encoder relies on a single-layer graph neural network to capture node representations. While effective for single-layer problems, it does not account for interactions between network layers. On the other hand, MultiDismantler introduces the Multiplex Graph Neural Network (MGNN), which integrates intra-layer graph convolution for local feature aggregation and a node-level inter-layer attention mechanism to learn dependencies across layers. This dual mechanism enables MultiDismantler to capture both intra-layer structures and inter-layer interactions, making it uniquely suited for dismantling multi-layer networks.

\textbf{Fusion of node scores across layers}.
FINDER is designed for the dismantling of single-layer network. In its generalization to the dismantling of interdependent networks,
nodes' scores are computed independently for each network layer meaning that the dismantling strategy is conceived from a single-layer perspective. This approach does not consider the interplay between layers in interdependent networks. MultiDismantler addresses this limitation by using a layer-level fusion mechanism to dynamically fuse node scores across layers. This fusion ensures that the final dismantling strategy accounts for both intra-layer importance and cross-layer interdependencies, a critical feature absent in FINDER.

\subsection{Similarities}
FINDER employs three key techniques to improve dismantling performance, which are also adopted and extended in MultiDismantler.

\textbf{Virtual node representation}. FINDER introduces virtual nodes to encode the global state of the network. Each virtual node is connected to all residual nodes in a layer but does not act as a neighbor to other nodes. The virtual nodes could inherently capture more complex graph information than the traditional sum pooling method. Similarly, MultiDismantler introduces a virtual node in each layer of the interdependent network to represent the state of that specific layer.

\textbf{Reconstruction loss}. To preserve the original network topology during training, FINDER incorporates a reconstruction loss function. This ensures that embeddings of nodes reflect their structural and positional properties within the graph. MultiDismantler follows the same principle by applying reconstruction loss to multiplex networks. This is particularly crucial for interdependent networks, as it allows the embeddings to encode both intra-layer and inter-layer relationships accurately, maintaining the integrity of the representation as dismantling progresses.

\textbf{Cross product for state-action interaction}. FINDER uses the cross product operation to model dependencies between network states and node actions during the decoding process. Similarly, MultiDismantler adopts this approach to capture the relationship between network states and node removal actions in interdependent networks.

\subsection{Experimental Comparison}
To ensure a fair comparison between FINDER and MultiDismantler, we modified FINDER to incorporate the same reward mechanism as of MultiDismantler. As shown in Supplementary Table \ref{table-as-reward}, FINDER still performs significantly worse than MultiDismantler. This finding demonstrates that the performance gap between the two methods is not solely due to the reward mechanism but arises also from the fundamental architectural differences and innovations introduced in MultiDismantler. Specifically, the absence of the GMM-generated training data and the inter-layer attention mechanism in FINDER makes it less capable of capturing the complex dependencies in interdependent networks.

\begin{table}[H]
\caption{Performance comparison of FINDER and MultiDismantler using the same reward mechanism. The best results are highlighted with bold fonts (corresponding to the lowest AUDC values).}
\centering
\resizebox{\textwidth}{!}{
\begin{tabular}{c|ccccccccc}
\hline
AUDC & USAir & FAO & Celegans & Drosophila & Fb\&Tw & NetSci & Sacchpomb & Homo & Sanremo2016 \\ 
\hline
FINDER & 0.077 & 0.281 & 0.243 & 0.115 & 0.344 & 0.311 & 0.060 & 0.043 & 0.003 \\

MultiDismantler & \textbf{0.053} & \textbf{0.252} & \textbf{0.162} & \textbf{0.056} &\textbf{0.319} & \textbf{0.051} & \textbf{0.049} & \textbf{0.027} & \textbf{0.001} \\
\hline
\end{tabular}
}
\label{table-as-reward}
\end{table}

\section{Experimental Details}
\subsection{Dataset statistics}
\label{Dataset statistics}
We provide details on the real-world datasets we analyzed. All networks are considered undirected and unweighted in our study. Some of the interdependent networks we analyzed contain more than two layers. Unless specified otherwise, in our analysis, we selected the two layers with largest number of edges. Also, in the analysis of single-layer networks, we focused our attention on the layer with the largest number of edges.

\textbf{Us air transportation}~\cite{radicchi2015percolation} is obtained by considering flights operated by Delta Airlines, American Airlines
and United Airlines in January 2014; a network layer represents one air carrier. In the network,  airports are nodes, and connections represent the existence of at least a flight between the two airports.

\textbf{FAO multiplex trade network}~\cite{de2015structural} is an economic network in which layers represent products, nodes are countries, and edges at each layer represent import/export relationships of a specific food product among countries. Data can be obtained from the Food and Agriculture Organization (FAO) and correspond to trading of year 2010.

\textbf{Celegans multiplex GPI network, Drosophila multiplex GPI network, Sacchpomb multiplex GPI network and Homo multiplex GPI network}~\cite{stark2006biogrid} are biological networks in which layers represent different types of genetic and protein interactions, nodes are genes or proteins, and edges at each layer represent specific interactions between these entities.The data for these multiplex networks is sourced from the BioGRID database (\url{thebiogrid.org}), which archives genetic and protein interaction data from humans and various model organisms.

\textbf{Facebook\&Twitter}~\cite{cao2016bass} is a multiplex network constructed from two real-world social networks as collected and published by Cao and Yong. Each user is a node, and connections are social network contacts on Facebook and Twitter.

\textbf{NetSci co-authorship}~\cite{de2015identifying} is a collaboration network where nodes represent scientists; two nodes are connected if the corresponding authors co-authored a paper. Layers are given by arXiv categories. To restrict the analysis to a well-defined topic of research, only papers with ``networks'' in the title or abstract were considered. The dataset contains all papers appeared on \url{arXiv.org} until May 2014.

\textbf{Sanremo2016}~\cite{de2020unraveling} is a social multiplex network in which layers represent different types of interactions among users on Twitter during the final of the Sanremo Music Festival in 2016. The dataset consists of three distinct layers: retweets, mentions, and replies, capturing the various forms of user engagement and communication during this exceptional event.

\begin{table}[H]
\caption{Statistical properties of real-world networks}
\centering
\begin{tabular}{cccccc}
\hline
Dataset & Category & Layers & Nodes & Edges & The two layers with the largest number of edges \\
\hline
UsAir & Transport & 3 & 84 & 1885 & Am. Air.-Delta \\
FAO & Financial & 364 & 214 & 318346 & Food\_prep-Crude 
\\
Celegans & Genetic & 3 & 279 & 5863 & Physical-Add. Gen. \\

Drosophila & Genetic & 3 & 676 & 6144 & Supp. Gen.-Add. Gen. \\ 

Fb\&Tw & Social & 2 & 1043 & 9594 & Facebook-Twitter \\
NetSci & Social & 3 & 1400 & 11173 & Data\_an-Dis\_nn \\

Sacchpomb & Genetic & 7 & 4092 & 63676 & Supp. Gen.-Add. Gen. \\
Homo & Genetic & 7 & 18222 & 170899 &  Direct-Physical \\
Sanremo2016 & Social & 3 & 56562 & 461838 & RT-MT\\
\hline
\end{tabular}
\label{real-network-properties}
\end{table}

\subsection{Model Parameters}
\begin{table}[H]
\caption{Hyper-parameters setting}
\centering
\begin{tabular}{ccc}
\hline
Model & Hyper-parameter & Value \\
\hline
\multirow{5}{*}{GMM}
& inter-layer degree correlation ($d$) & 0.2 \\
& inter-layer similarity correlation ($g$) & 0.5 \\
& power-law degree distribution exponent ($\gamma$) & 2.5 \\
& expected mean degree ($\overline{k}$) & 6.0 \\
&the network clustering control parameter ($T$)& 0.4\\
\hline
\multirow{10}{*}{MultiDismantler}
& discount factor & 1.0 \\
& update time of targeted network & $10^3$ \\
& embedding size & 64 \\
& neighbor-aggregation iterations & 3 \\
& max iterations of training & $10^5$ \\
& the learning rate used by Adam optimizer & $10^{-4}$\\
& experience replay buffer size & $10^5$ \\
& weight of reconstruction loss & $10^{-3}$ \\
& steps for multi-step Q-learning algorithm & 5 \\
& mini-batch training samples & 64 \\
\hline
\end{tabular}
\label{parameter-setting}
\end{table}

\section{More Experiments}
\subsection{Ablation study on factors influencing MultiDismantler's performance}
\begin{table}[H]
\caption{Ablation study. We conduct three ablation studies to assess the contributions of essential components. The best results are highlighted with bold fonts (corresponding to the lowest AUDC values).}
\centering
\subfloat[Ablation study on inter-layer attention]{
\begin{tabular}{c|ccccccccc}
\hline
AUDC & USAir & FAO & Celegans & Drosophila & Fb\&Tw & NetSci & Sacchpomb & Homo & Sanremo2016 \\ 
\hline
w/o inter-layer attention & 0.062 & 0.255 & 0.214 & 0.057 & 0.332 & 0.056 & 0.050 & 0.028 & \textbf{0.001} \\

w inter-layer attention & \textbf{0.053} & \textbf{0.252} & \textbf{0.162} & \textbf{0.056} &\textbf{0.319} & \textbf{0.051} & \textbf{0.049} & \textbf{0.027} & \textbf{0.001} \\
\hline
\end{tabular}
\label{table-as-attention}
}

\vspace*{0.02\linewidth}
\subfloat[Ablation study on training corpus]{
\begin{tabular}{c|ccccccccc}
\hline
AUDC & USAir & FAO & Celegans & Drosophila & Fb\&Tw & NetSci & Sacchpomb & Homo & Sanremo2016 \\ 
\hline
BA & 0.054 & 0.432 & 0.203 & 0.059 & 0.343 & 0.069 & 0.057 & 0.037 & 0.002 \\

GMM & \textbf{0.053} & \textbf{0.252} & \textbf{0.162} & \textbf{0.056} &\textbf{0.319} & \textbf{0.051} & \textbf{0.049} & \textbf{0.027} & \textbf{0.001} \\

\hline
\end{tabular}
\label{table-as-GMM}
}

\vspace*{0.02\linewidth}

\subfloat[Ablation study on inter-layer similarity]{
\begin{tabular}{c|ccccccccc}
\hline
AUDC & USAir & FAO & Celegans & Drosophila & Fb\&Tw & NetsSci & Sacchpomb & Homo & Sanremo2016 \\ 
\hline
$g=0$ & \textbf{0.053} & 0.293 & 0.216 & 0.057 & 0.404 & 0.103 & 0.057 & 0.034 & 0.002 \\

$g=0.2$ & \textbf{0.053} & 0.261 & 0.177 & \textbf{0.053} & 0.363 & 0.071 & \textbf{0.048} & 0.028 & 0.002 \\

$g=0.5$ & \textbf{0.053} & \textbf{0.252} & \textbf{0.162} & 0.056 &\textbf{0.319} & 0.051 & 0.049 & \textbf{0.027} & \textbf{0.001} \\

$g=0.8$ & 0.055 & 0.260 & \textbf{0.162} &0.056 & 0.330 & 0.051 & 0.050 & 0.028 & 0.002 \\

$g=1.0$ & 0.055 & 0.318 & 0.163 & 0.059 & 0.330 & \textbf{0.048} & 0.051 & 0.029 & 0.002 \\

\hline
\end{tabular}
\label{table-as-g}
}
\label{table-as}
\end{table}

\subsection{Performance on other layers in real-world multiplex networks}
\begin{table}[H]
\caption{Dismantling performance on the other layers of real-world multiplex networks with unit costs. We selected the two layers with relatively more edges and the two layers with relatively fewer edges in each network. The best results are highlighted with bold fonts (corresponding the lowest AUDC values). }
\centering
\begin{tabular}{cccccccccc}
\hline
Dataset & Layer & HDA & CI & MinSum & FINDER & NIRM & CoreHLDA &EMD& MultiDismantler \\ 
\hline
\multirow{2}{*}{USAir}
&Am. Air.-United & 0.063 & 0.064 & 0.084 & 0.053 & 0.060& 0.064&\textbf{0.050}& 0.061 \\
&Delta-United & 0.074 & 0.076 & 0.178 & \textbf{0.048} & 0.076  & 0.069&0.056& \textbf{0.048} \\
\hline

\multirow{2}{*}{FAO}
&Fat\_nes-Bacon.   & 0.073 & 0.079 & 0.099 & 0.071 & 0.072  & 0.073&0.073& \textbf{0.068} \\
&Meat\_goat-Rye & \textbf{0.019} & 0.021 & 0.036 & 0.022 & \textbf{0.019}  & 0.020&0.020& \textbf{0.019} \\
\hline

\multirow{2}{*}{Celegans}
&Direct-Add. Gen. & 0.112 & 0.111 & 0.157 & 0.076 & 0.089  & 0.107&0.095& \textbf{0.075} \\
&Direct-Physical & 0.089 & 0.091 & 0.181 & \textbf{0.064}& 0.077  & 0.076&0.082& 0.068 \\
\hline

\multirow{2}{*}{Drosophila}
&Direct-Supp. Gen. & 0.010 & 0.012 & 0.021 & 0.013 & 0.010& 0.011&\textbf{0.009}& 0.011 \\
&Direct-Add. Gen. & 0.004 & \textbf{0.003} & 0.014 & \textbf{0.003} & 0.004  & 0.004&0.004& \textbf{0.003} \\
\hline

\multirow{2}{*}{NetSci}
&Data\_an-Stat\_mech & 0.025 & 0.029 & 0.105 & 0.012 & 0.022& 0.024&\textbf{0.011}& 0.027 \\
&Dis\_nn-Stat\_mech & 0.026 & 0.033 & 0.097 & 0.009 & 0.022& 0.028&\textbf{0.008}& 0.025 \\
\hline

\multirow{2}{*}{Sacchpomb}
&Physical-Add. Gen. & 0.006 & 0.013 & 0.055 & \textbf{0.003} & 0.004  & 0.005&0.004& \textbf{0.003} \\
&Physical-Supp. Gen. & 0.007 & 0.010 & 0.036 & \textbf{0.004} & 0.007  & 0.006&0.005& \textbf{0.004} \\
\hline

\multirow{2}{*}{Homo}
&Physical-Colocalization & 0.010 & 0.011 & 0.043 & 0.016& 0.009  & 0.009&0.008& \textbf{0.007} \\
&Direct-Colocalization & 0.010 & 0.011 & 0.029 &0.014 & 0.009  & 0.009&0.008& \textbf{0.007} \\
\hline

\multirow{2}{*}{Sanremo2016}
&RT-RE & \textbf{0.001} & \textbf{0.001} & 0.004 & \textbf{0.001} & \textbf{0.001}  & \textbf{0.001}&\textbf{0.001}& \textbf{0.001
}\\
&MT-RE & \textbf{0.001} & 0.002 & 0.003 & \textbf{0.001} & \textbf{0.001}  & 0.002&\textbf{0.001}& \textbf{0.001} \\
\hline

\hline
\end{tabular}
\label{table-real-otherlayer}
\end{table}

\subsection{Sequence consistency between MultiDismanlter and the optimal dismantling sequence}
\begin{table}[H]
\caption{Comparison with brute-force method for dismantling multiplex networks. All solutions show the first n nodes, where n is the length of the optimal sequence.
The best results are highlighted with bold fonts (corresponding the lowest AUDC values and the highest 1 - NED values).}
\centering
\subfloat[Dismantling Sequence of different algorithms and the optimal sequence identified with brute-force method.]{
\begin{tabular}{cccccc}
\hline
Algorithm & Synthetic-10 & Synthetic-15 & Synthetic-20 & Synthetic-25 & Synthetic-30 \\
\hline 
HDA &[7 4 6]& [11 14 2 13 4 6] &[19 10 2 12 0] &[18 19 13 7 11 24]& [13 5 24 20]\\
CI &[7 4 2] &[11 14 2 4 10 0] &[19 10 2 12 0] &[18 19 13 7 21 15] &[17 13 5 24]\\
MinSum &[1 5 0] & [11 6 9 13 14 2]&[16 19 2 6 7] &[18 19 7 2 21 13] &[24 13 0 3] \\
FINDER & [4 3 2]&[11 2 12 13 14 6] & [19 21 8 14 2]&[18 13 19 7 24 21] & [13 5 24 6]\\
NIRM  & [4 6 3]& [11 14 2 13 6 10] & [19 0 2 15 7] & [18 13 19 22 7 3] & [13 5 24 6]\\
 CoreHLDA& [7 4 0]& [11 14 13 4 9 10]& [19 12 0 17 15]& [18 7 24 13 12 22]&[13 5 24 6]\\
 EMD& [4 7 6]& [11 14 13 2 6 12]& [19 14 2 0 17]& [18 7 24 19 22 13]&[13 24 2 17]\\
MultiDismantler &[4 7 6] & [11 14 13 2 6 10] & [19 0 2 17 15] & [18 13 19 7 22 24] &[13 24 5 6] \\ 
\hline
\multirow{4}{*}{Optimal}&[4 7 6] & [11 14 13 2 6 10] & [19 0 2 7 15]& [18 13 19 7 22 3] & [13 24 5 6]
\\&[4 7 9]&[11 14 13 2 6 12]&[19 0 2 17 15]&[18 13 19 7 22 24]&[13 24 5 20]
\\&&[14 11 13 2 6 10]&[19 0 7 2 15]
\\&&[14 11 13 2 6 12]&[19 0 17 2 15]&&\\
\hline
\end{tabular}
\label{table-bf-solution}
}

\vspace{0.02\linewidth}

\subfloat[Dismantling sequence similarity and dismantling performance comparison of different algorithms]{
\begin{tabular}{cccccccccc}
\hline
Dataname & Metric &HDA & CI & MinSum & FINDER & NIRM  & CoreHLDA&EMD& MultiDismantler\\
\hline
\multirow{2}{*}{Synthetic-10}
&1 - NED&0.333 & 0.000 & 0.000 & 0.333 & 0.333  & 0.000&\textbf{1.000}& \textbf{1.000}\\
& AUDC & 0.130 & 0.220 & 0.229 & 0.100 &0.110  & 0.140&\textbf{0.080}&\textbf{0.080}\\
\hline
\multirow{2}{*}{Synthetic-15}
&1 - NED&0.500&0.500&0.167&0.333&0.667 & 0.667&\textbf{1.000}&\textbf{1.000}\\
&AUDC&0.200&0.262 &0.284 &0.217 &0.187  & 0.201&\textbf{0.178}&\textbf{0.178}\\
\hline
\multirow{2}{*}{Synthetic-20}
&1 - NED&0.400&0.400&0.200&0.200&0.600 & 0.600&0.400&\textbf{1.000}\\
&AUDC&0.178 &0.183 &0.275 &0.138 &0.075  & 0.108&0.085&\textbf{0.073}\\
\hline
\multirow{2}{*}{Synthetic-25}
&1 - NED&0.500 & 0.333 & 0.333 & 0.667 &0.667  & 0.167&0.333& \textbf{1.000}\\
&AUDC&0.053 &0.093 &0.085 & 0.050 &0.048  & 0.077&0.072&\textbf{0.047}\\
\hline
\multirow{2}{*}{Synthetic-30}
&1 - NED&0.500 & 0.250 & 0.000 & 0.500 & 0.500  & 0.500&0.500& \textbf{1.000} \\
&AUDC&0.031 &0.091 &0.083 &0.030 &0.031  & 0.031&0.027& \textbf{0.024}\\
\hline
\end{tabular}
\label{table-bf-metric}
}
\label{table-bf}
\end{table}

\begin{figure}[H]
\centering
\begin{subfigure}{0.3\linewidth}
    \centering
    \includegraphics[width=\linewidth]{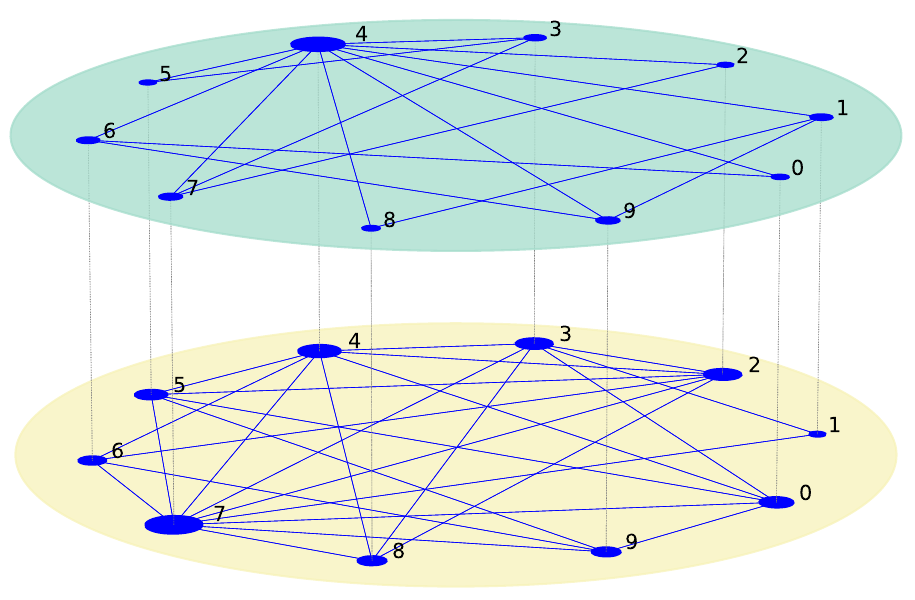}
    \subcaption{} 
    \label{baoli_Synthetic10}
\end{subfigure}
\hfill
\begin{subfigure}{0.3\linewidth}
    \centering
    \includegraphics[width=\linewidth]{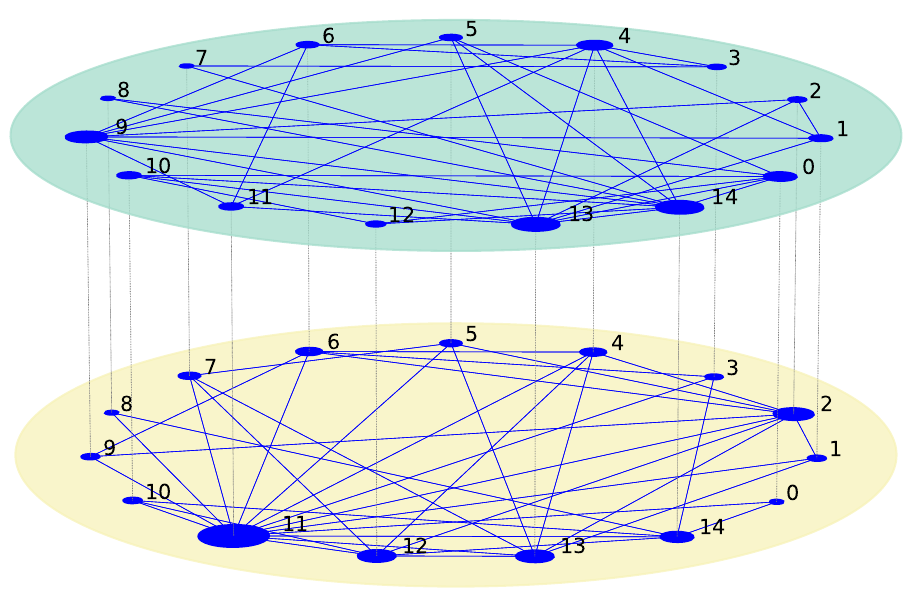}
    \subcaption{} 
    \label{baoli_Synthetic15}
\end{subfigure}
\hfill
\begin{subfigure}{0.3\linewidth}
    \centering
    \includegraphics[width=\linewidth]{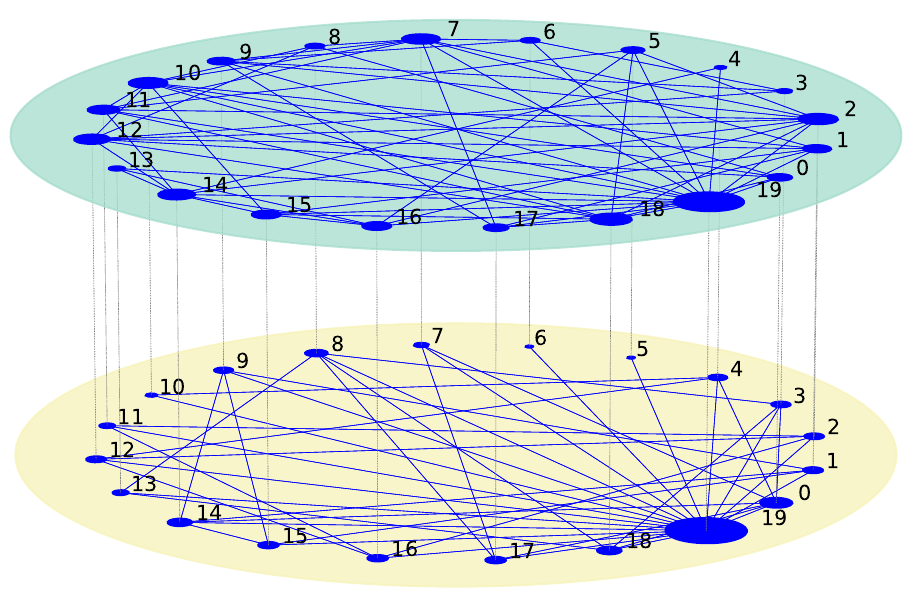}
    \subcaption{} 
    \label{baoli_Synthetic20}
\end{subfigure}

\begin{subfigure}{0.3\linewidth}
    \centering
    \includegraphics[width=\linewidth]{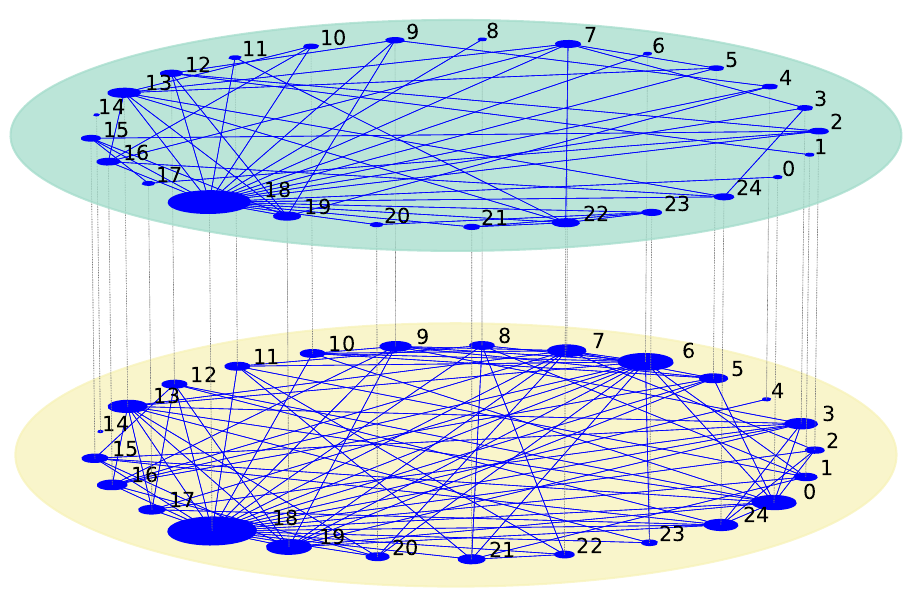}
    \subcaption{}
    \label{baoli_Synthetic25}
\end{subfigure}
\hspace{0.1\linewidth}
\begin{subfigure}{0.3\linewidth}
    \centering
    \includegraphics[width=\linewidth]{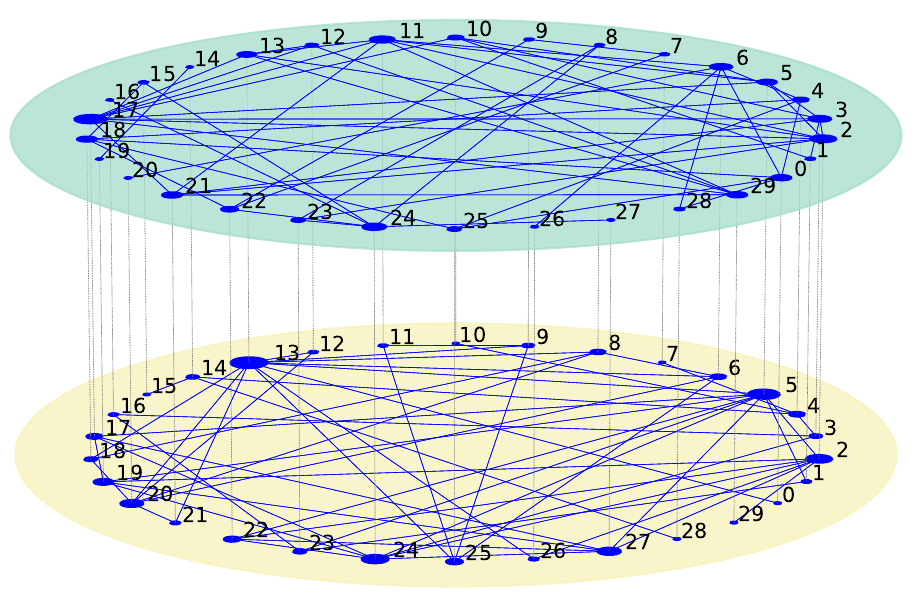}
    \subcaption{} 
    \label{baoli_Synthetic30}
\end{subfigure}

\caption{Interdependent network topology used in the experiment on comparing with brute-force method for identifying the optimal dismantling sequence. Networks \textbf{(a)}, \textbf{(b)}, \textbf{(c)}, \textbf{(d)}, and \textbf{(e)} represent synthetic networks with 10, 15, 20, 25, and 30 nodes, respectively.}
\label{SIFigure:BruteForceTopology}
\end{figure}

\subsection{Dismantling performance under more cost scenarios of real-world networks}
\begin{table}[H]
\caption{Dismantling performance on real-world multiplex networks.  We choose the the two layers with the largest number of edges, and dismantle the layers under Random, Normal, and Poisson costs. The best results are highlighted with bold fonts (corresponding to the lowest AUDC values).}
\centering
\subfloat[Random costs]{
    \begin{tabular}{ccccccccc}
    \hline
    Dataset  & HDA & CI & MinSum & FINDER & NIRM  & CoreHLDA & EMD & MultiDismantler \\ 
    \hline
    UsAir
    & 0.079 & 0.100 & 0.173 & 0.138 & 0.060  & 0.060 & 0.060& \textbf{0.057} \\
   FAO &  0.273 & 0.277 & 0.293 & 0.361 & 0.266  & 0.256&0.253& \textbf{0.251} \\
   Celegans & 0.183 & 0.198 & 0.323 & 0.226 & 0.163  & 0.181&0.167& \textbf{0.142} \\
  Drosophila   & 0.070 & 0.075 & 0.083 & 0.117 & 0.067  & 0.059&\textbf{0.055}& 0.057\\
   Fb\&Tw  & 0.339 & 0.343 & 0.342 & 0.378 & 0.338  & 0.343&0.334& \textbf{0.288} \\
  NetSci & 0.050 & 0.054 & 0.116 & 0.122& \textbf{0.048}  & 0.052&0.054& 0.053 \\
   Sacchpomb & 0.055 & 0.060 & 0.076 & 0.110 & 0.053  & 0.055&0.053& \textbf{0.047} \\
   Homo & 0.031 & 0.032 & 0.055 & 0.043 & 0.030  & 0.030&0.028& \textbf{0.027} \\
    Sanremo2016 & \textbf{0.002} & 0.003 & 0.004 & 0.015 & \textbf{0.002}  & \textbf{0.002}&\textbf{0.002}& \textbf{0.002} \\
    \hline
    \end{tabular}
    \label{table-real-weighted}
}   
\vspace*{0.02\linewidth}
\subfloat[Normal costs]{
\begin{tabular}{ccccccccc}
\hline
Dataset & HDA & CI & MinSum & FINDER & NIRM  & CoreHLDA&EMD& MultiDismantler \\ 
\hline
UsAir & 0.070 & 0.088 & 0.124 & 0.120 & \textbf{0.054}  & 0.072&0.059& 0.056 \\

FAO  & 0.264 & 0.271 & 0.283 & 0.327 & 0.264  & 0.256&\textbf{0.250}& 0.253\\

Celegans & 0.189 & 0.208 & 0.305 & 0.329 & \textbf{0.163}  & 0.183&0.168& 0.164 \\

Drosophila & 0.065 & 0.071 & 0.083 & 0.074 & 0.063  & 0.062&0.059& \textbf{0.058}\\

Fb\&Tw  & 0.339 & 0.341 & 0.348 & 0.504& 0.337  & 0.342&0.331& \textbf{0.324} \\

NetSci & 0.051 & 0.055 & 0.117 & 0.108 & \textbf{0.049}  & \textbf{0.049}&0.053& 0.053\\
Sacchpomb & 0.056 & 0.060 & 0.080 & 0.071& 0.054  & 0.555&0.052& \textbf{0.050} \\
Homo & 0.030 & 0.032 & 0.054 & 0.112& 0.030  & 0.029&\textbf{0.028}& \textbf{0.028} \\
Sanremo2016 & \textbf{0.002} & 0.003 & 0.004 & 0.030 & \textbf{0.002}  & \textbf{0.002}&\textbf{0.002}& \textbf{0.002} \\
\hline
\end{tabular}
\label{table-real-normal-weighted}}

\vspace*{0.02\linewidth}

\subfloat[Poisson costs]{
\begin{tabular}{ccccccccc}
\hline
Dataset & HDA & CI & MinSum & FINDER & NIRM  &CoreHLDA&EMD& MultiDismantler\\ 
\hline
UsAir & 0.070 & 0.088 & 0.124 &0.140  & 0.067  & 0.070&\textbf{0.050}& 0.054\\

FAO  & 0.267 & 0.283 & 0.283 & 0.326 & 0.267  & 0.256&0.252& \textbf{0.240} \\

Celegans & 0.190 & 0.206 & 0.300 & 0.336 & \textbf{0.162}  & 0.187&0.167& 0.168 \\

Drosophila & 0.065 & 0.071 & 0.088 & 0.099 & \textbf{0.059}  & 0.060&0.062& \textbf{0.059}\\

Fb\&Tw  & 0.344 & 0.335 & 0.345 & 0.382 & 0.338  & 0.344&0.332& \textbf{0.310} \\

NetSci & 0.050 & 0.055 & 0.118 & 0.125 & \textbf{0.048}  & 0.052&0.052& 0.050\\

Sacchpomb & 0.055 & 0.059 & 0.078 & 0.119 & 0.055  & 0.050&0.052& \textbf{0.049} \\
Homo & 0.029 & 0.032 & 0.054 & 0.045 & 0.030  & 0.030&\textbf{0.027}& \textbf{0.027} \\
Sanremo2016 & \textbf{0.002} & 0.003 & 0.004 & 0.006 & \textbf{0.002}  & \textbf{0.002}&\textbf{0.002}& \textbf{0.002} \\
\hline
\end{tabular}
\label{table-real-poisson-weighted}
}

\label{table-real-inductive}
\end{table}

\subsection{Dismantling performance on single-layer networks}
\label{single-layer-ND}

We applied MultiDismantler to the single-layer network dismantling task. We selected the layer with the most edges from the real-world networks to perform single-layer
network dismantling under the unit cost scenario. Specifically, we model the single-layer network as a two-layer interdependent network with identical topologies. In this configuration,
cascading failures do not occur, which is equivalent to directly dismantling the single-layer
network. The original well-trained MultiDismantler is referred to as MD. Table \ref{tale-single} shows that MultiDismantler, trained specifically for dismantling multiplex networks, also outperforms all comparison algorithms in single-layer dismantling tasks, demonstrating its versatility and generalization capabilities.

To further investigate the key components contributing to MultiDismantler's performance, we trained two variants. In the first variant, we removed the inter-layer message-passing mechanism from the encoding architecture, referring to it as ``w/o inter-layer attention." As shown in Table \ref{tale-single}, this variant exhibits overall inferior performance compared to the original MD, emphasizing the importance of inter-layer attention in enhancing MultiDismantler's dismantling performance. 

In the second variant, we retrained MultiDismantler with the training corpus generated with BA networks instead of GMM, while keeping all other settings unchanged. This variant, referred to as ``MD w/o GMM" shows a significant drop in dismantling performance, as indicated in the last column of Table \ref{tale-single}. 

Overall, these experiments clarify the contributions of the inter-layer attention mechanism and the importance of GMM-generated training data to the performance of MultiDismanter. Both the inter-layer attention mechanism and the GMM-generated multiplex networks, which include inter-layer geometric correlations, are integral components of the inter-layer framework. These findings indicate that the superior performance of our method primarily stems from inter-layer components, as they enable the model to effectively exploit the structural complexity and dependencies unique to multiplex networks.

 



\begin{table}[H]
\centering
\caption{Performance comparison of dismantling algorithms on singer layer. The best results are highlighted with bold fonts (corresponding to the lowest AUDC values).}
\resizebox{\textwidth}{!}{
\begin{tabular}{lcccccccc}
\hline
Dataset & HDA & CI & MinSum & FINDER & NIRM & MD & MD (w/o inter-layer attention) & MD (w/o GMM)
 \\ 
\hline
USAir      & 0.150 & 0.149 & 0.209 & 0.148 & 0.145 & 0.140 & \textbf{0.139} & 0.140\\
FAO        & 0.323 & 0.327 & 0.327 & 0.318 & 0.323 & \textbf{0.305} & 0.310 & 0.409 \\
Celegans   & 0.278 & 0.281 & 0.303 & 0.276 & 0.286 & \textbf{0.267} & 0.289 & 0.301 \\
Drosophila & 0.100 & 0.099 & 0.108 & 0.101 & 0.104 & \textbf{0.095} & 0.096 & 0.101 \\
Fb\&Tw     & 0.336 & 0.342 & 0.381 & 0.331 & 0.332 & \textbf{0.320} & 0.336 & 0.345 \\
NetSci     & 0.095 & 0.097 & 0.179 & 0.095 & 0.095 & \textbf{0.093} & 0.100 & 0.109 \\
Sacchpomb  & 0.083 & 0.086 & 0.216 & 0.080 & 0.083 & \textbf{0.079} & 0.082 & 0.227 \\
Homo       & 0.037 & 0.096 & 0.075 & 0.036 & 0.037 & \textbf{0.035} & 0.036 & 0.044 \\
Sanremo2016 & 0.025 & 0.026 & 0.122 & \textbf{0.024}& 0.025 & \textbf{0.024} & 0.025 & 0.034 \\
\hline
\end{tabular}
}
\label{tale-single}
\end{table}

\subsection{Dismantling performance of single-layer algorithms using aggregated scores}
\label{agg}
In this study, we applied single-layer algorithms to each layer individually and then selected the maximum dismantling
scores from each layer, sorting them to obtain a dismantling sequence of nodes. We also extended our experiments to include single-layer algorithms using aggregated scores under unit cost scenario. Specifically, we collapsed the multiplex networks into single-layer representations, where edges from all layers were aggregated without distinguishing their types. By applying single-layer dismantling algorithms to these aggregated networks, we generated dismantling sequences for comparison.

The results, shown in Supplementary Tables \ref{table-real-world-unit} and \ref{table-aggrgated}, demonstrate that MultiDismantler achieves overall best performance when compared against single-layer algorithms applied to aggregated scores and max scores. Notably, FINDER and NIRM, which rank as the second-best and third-best single-layer algorithms in Table \ref{table-aggrgated}, show no significant improvement under the aggregated approach. NIRM's performance declines in some datasets, such as USAir and Celegans.

\begin{table}[H]
\centering
\caption{Performance comparison of single-layer dismantling algorithms using aggregated scores. The best results are highlighted with bold fonts (corresponding to the lowest AUDC values).}
\begin{tabular}{lcccccc}
\hline
Dataset     & HDA    & CI     & FINDER & MinSum & NIRM   & MultiDismantler \\ 
\hline
USAir       & 0.070  & 0.071  & 0.084  & 0.194  & 0.070  & \textbf{0.053}  \\
FAO         & 0.256  & 0.257  & \textbf{0.251}  & 0.286  & 0.255  & 0.252  \\
Celegans      & 0.180  & 0.186  & 0.167  & 0.413  & 0.178  & \textbf{0.162}  \\
Drosophila          & 0.063  & 0.066  & 0.058  & 0.124  & 0.061  & \textbf{0.056}  \\
Fb\&Tw      & 0.340  & 0.343  & 0.331  & 0.365  & 0.334  & \textbf{0.319}  \\
NetSci         & 0.050  & 0.051  & \textbf{0.048}  & 0.191  & \textbf{0.048}  & 0.051  \\
Sacchpomb       & 0.054  & 0.056  &0.053  & 0.180  & 0.054  & \textbf{0.049}  \\
Homo        & 0.029  & 0.031  & 0.032  & 0.078  & 0.029  & \textbf{0.027}  \\
Sanremo2016 & 0.002  & 0.002  & 0.002  & 0.016  & 0.002  & \textbf{0.001}  \\
\hline
\end{tabular}
\label{table-aggrgated}
\end{table}

\section{Visualization for Dismantling Process}

\begin{figure}[H]
\centering
\begin{subfigure}{0.3\linewidth}
    \centering
    \includegraphics[width=\linewidth]{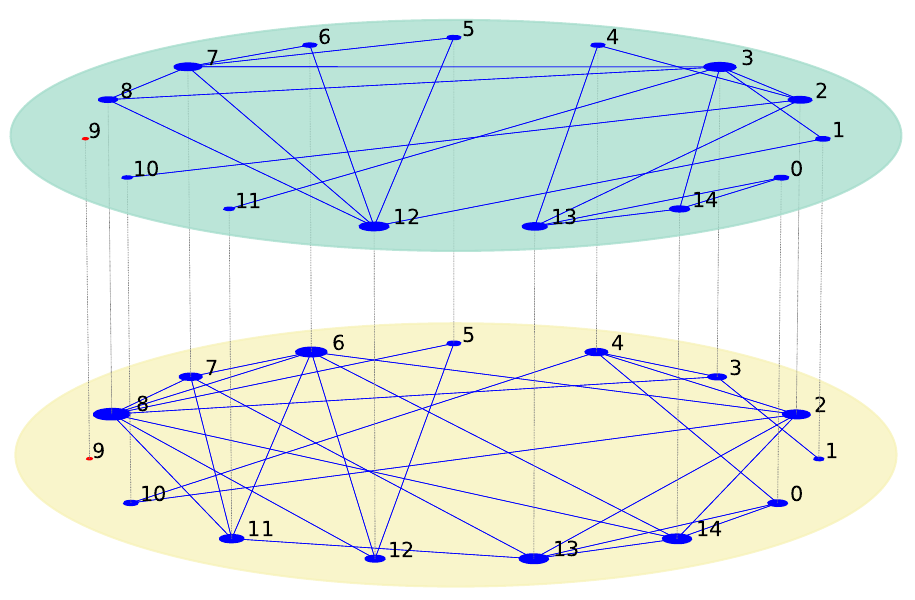}
    \subcaption{} 
    \label{Synthetic-15-HDA-new1}
\end{subfigure}
\hfill
\begin{subfigure}{0.3\linewidth}
    \centering
    \includegraphics[width=\linewidth]{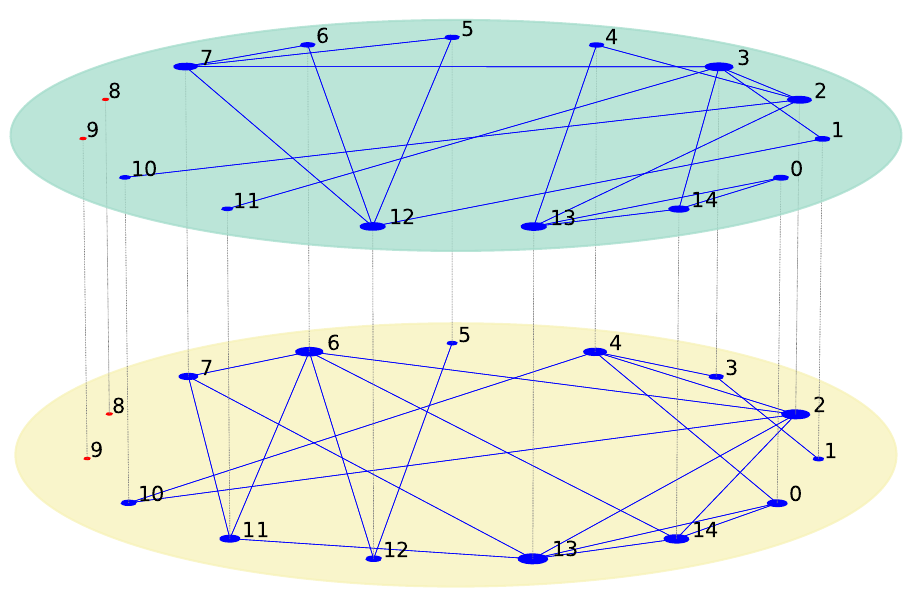}
    \subcaption{} 
    \label{Synthetic-15-HDA-new2}
\end{subfigure}
\hfill
\begin{subfigure}{0.3\linewidth}
    \centering
    \includegraphics[width=\linewidth]{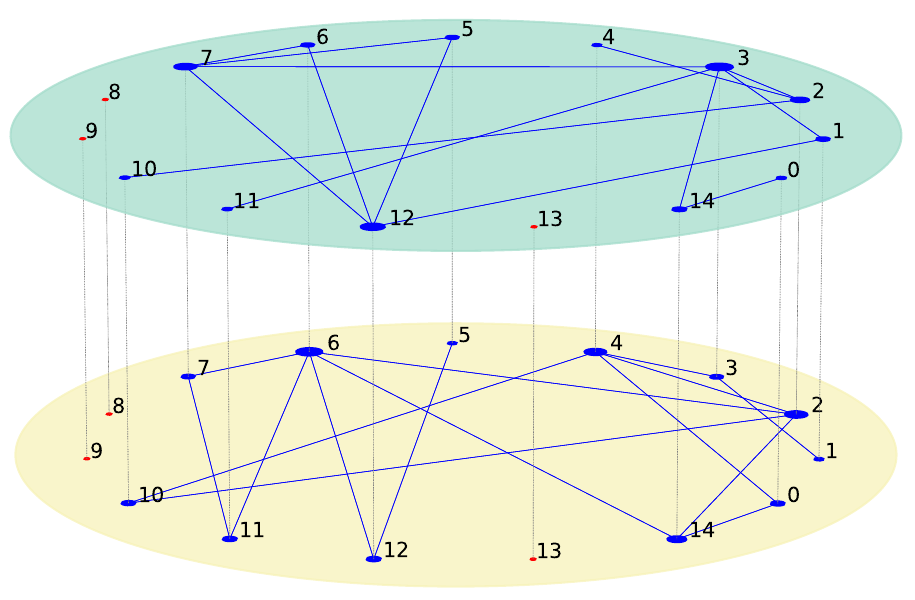}
    \subcaption{} 
    \label{Synthetic-15-HDA-new3}
\end{subfigure}

\vspace{0.5cm} 

\begin{subfigure}{0.3\linewidth}
    \centering
    \includegraphics[width=\linewidth]{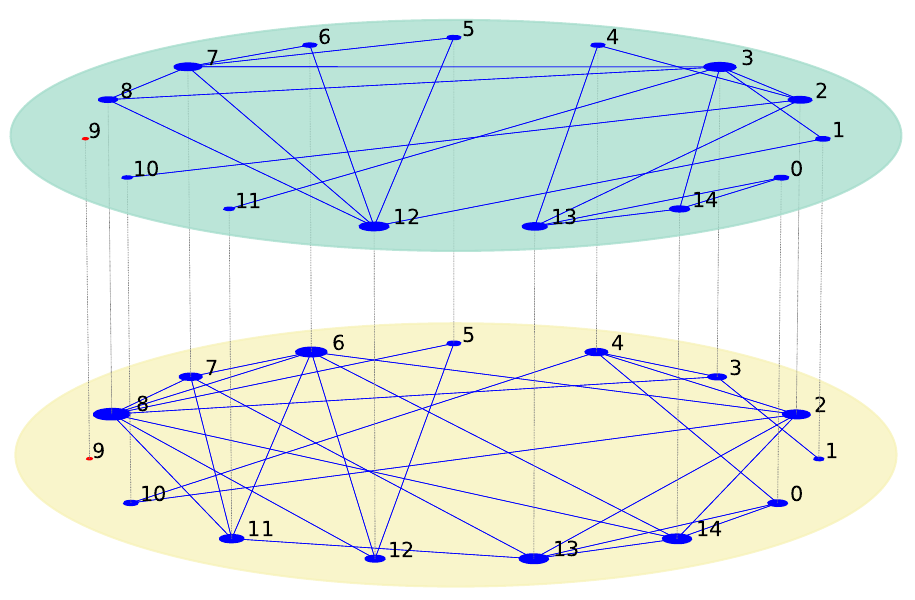}
    \subcaption{} 
    \label{Synthetic-15-OUR-new1}
\end{subfigure}
\hfill
\begin{subfigure}{0.3\linewidth}
    \centering
    \includegraphics[width=\linewidth]{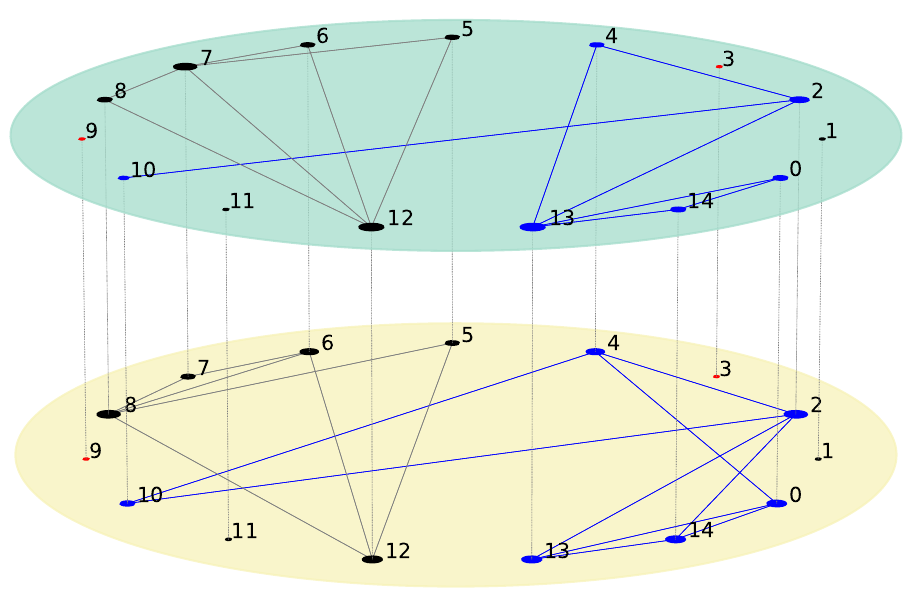}
    \subcaption{}
    \label{Synthetic-15-OUR-new2}
\end{subfigure}
\hfill
\begin{subfigure}{0.3\linewidth}
    \centering
    \includegraphics[width=\linewidth]{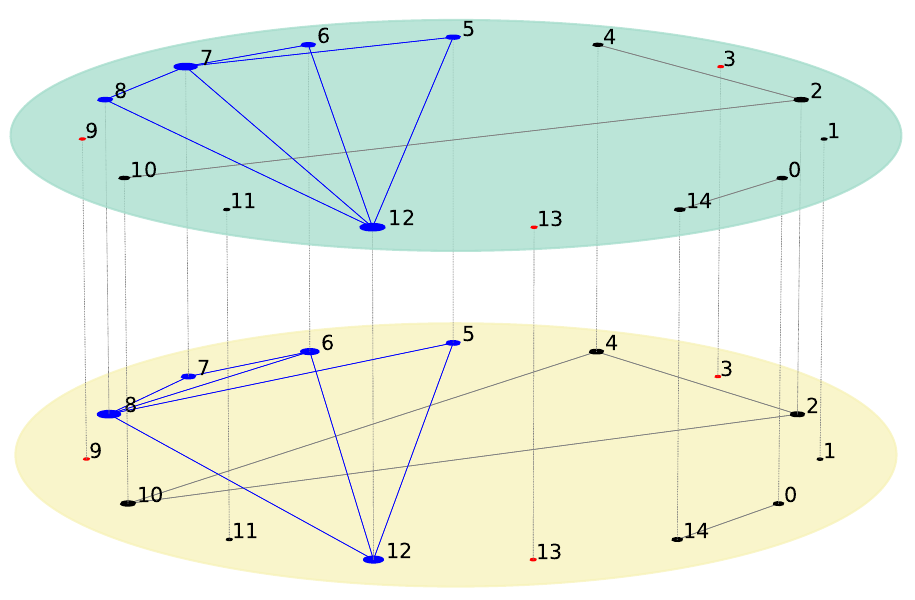}
    \subcaption{}
    \label{Synthetic-15-OUR-new3}
\end{subfigure}

\caption{The dismantling process of a synthetic interdependent network consisting of 15 nodes is illustrated in panels \textbf{(a)}, \textbf{(b)}, and \textbf{(c)} for the HDA algorithm, and in panels \textbf{(d)}, \textbf{(e)}, and \textbf{(f)} for the MultiDismantler algorithm. In these diagrams, blue nodes and edges represent the LMCC, while red nodes indicate the removed nodes. Initially, both HDA and MultiDismantler remove node 9. In the second step, MultiDismantler proceeds to remove the bridge node 3, as shown in panel \textbf{(e)}. This action divides the network into two similar parts, with one consisting of 5 nodes and the other consisting of 6 nodes. In contrast, the HDA algorithm selects node 8, which fails to substantially disrupt the network, leaving the LMCC comprising 13 nodes. Overall, MultiDismantler demonstrates a more effective disintegration strategy compared to the HDA algorithm.}
\label{fig-synthetic-15-dismantling}
\end{figure}

\begin{figure}[H]
\centering
\begin{subfigure}{0.3\linewidth}
    \centering
    \includegraphics[width=\linewidth]{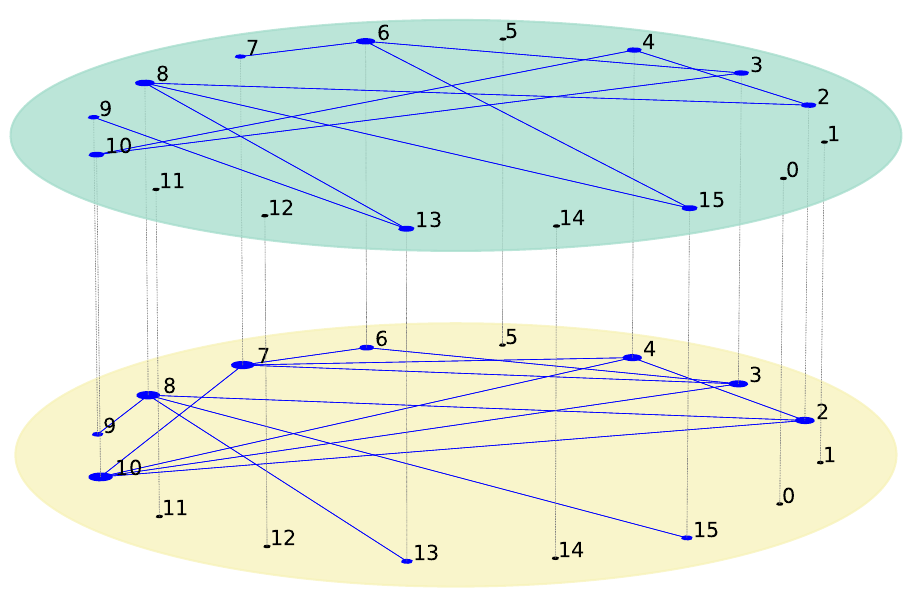} 
    \subcaption{} 
    \label{Flo_family_HDA_0}
\end{subfigure}
\hfill
\begin{subfigure}{0.3\linewidth}
    \centering
    \includegraphics[width=\linewidth]{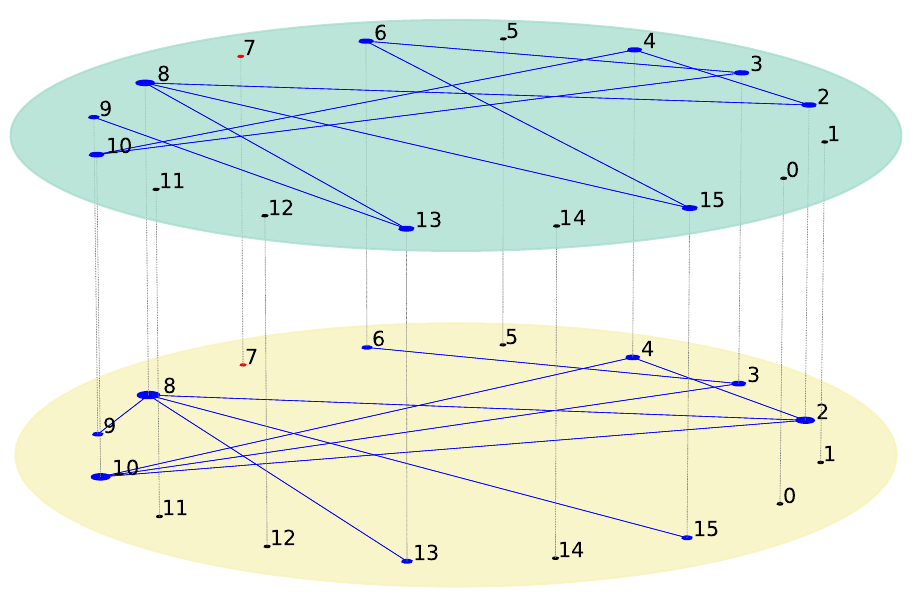}
    \subcaption{} 
    \label{Flo_family_HDA_1}
\end{subfigure}
\hfill
\begin{subfigure}{0.3\linewidth}
    \centering
    \includegraphics[width=\linewidth]{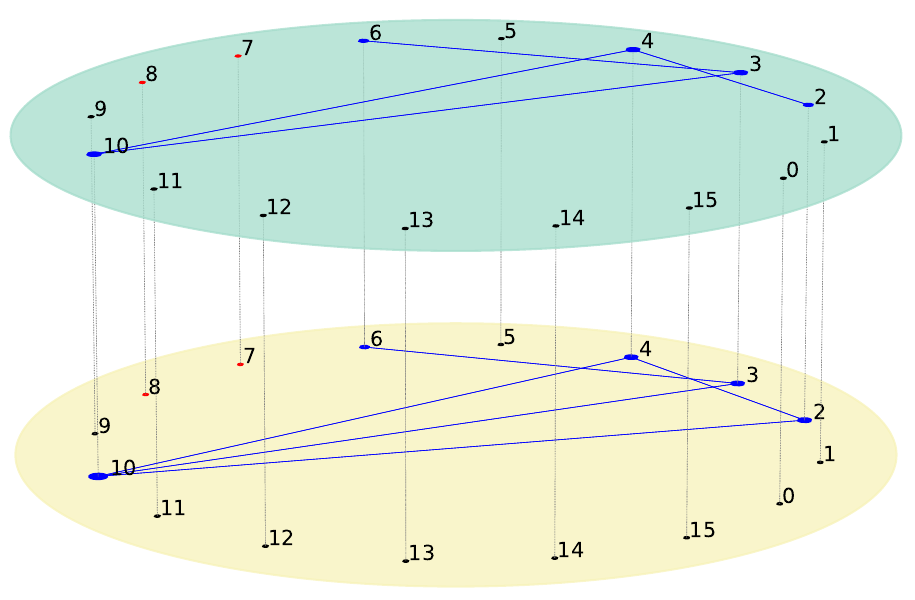}
    \subcaption{} 
    \label{Flo_family_HDA_2}
\end{subfigure}

\vspace{0.5cm} 

\begin{subfigure}{0.3\linewidth}
    \centering
    \includegraphics[width=\linewidth]{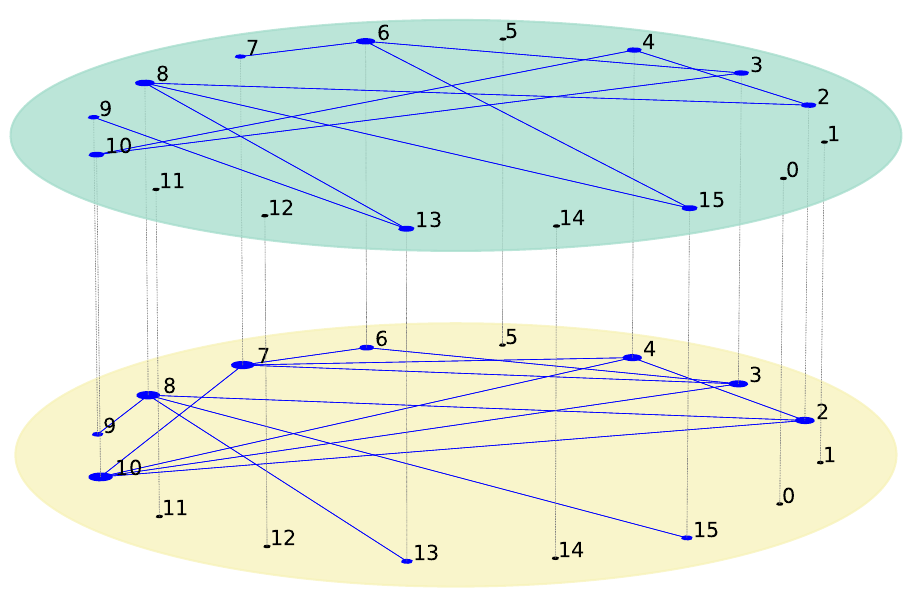}
    \subcaption{}
    \label{Flo_family_OUR_0}
\end{subfigure}
\hfill
\begin{subfigure}{0.3\linewidth}
    \centering 
    \includegraphics[width=\linewidth]{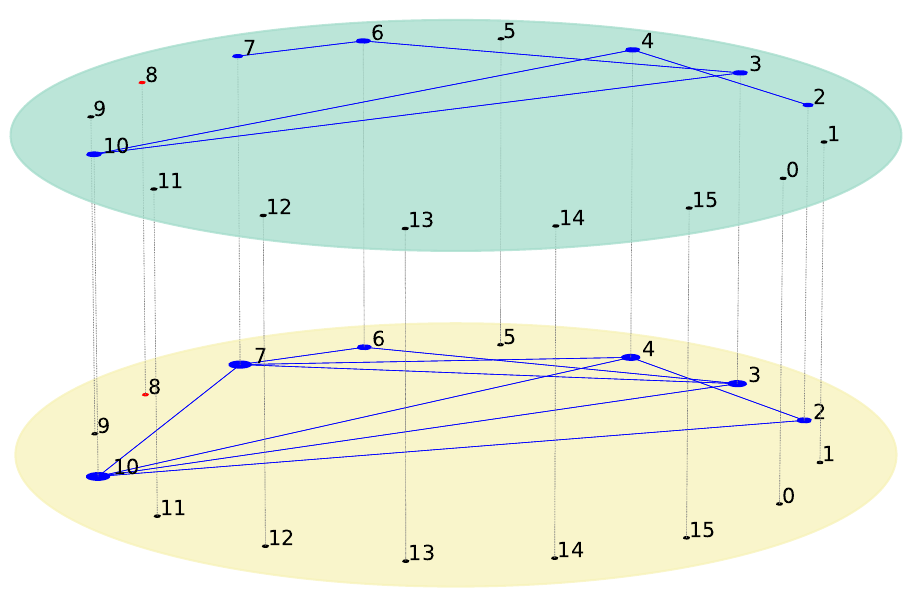}
    \subcaption{}
    \label{Flo_family_OUR_1}
\end{subfigure}
\hfill
\begin{subfigure}{0.3\linewidth}
    \centering
    \includegraphics[width=\linewidth]{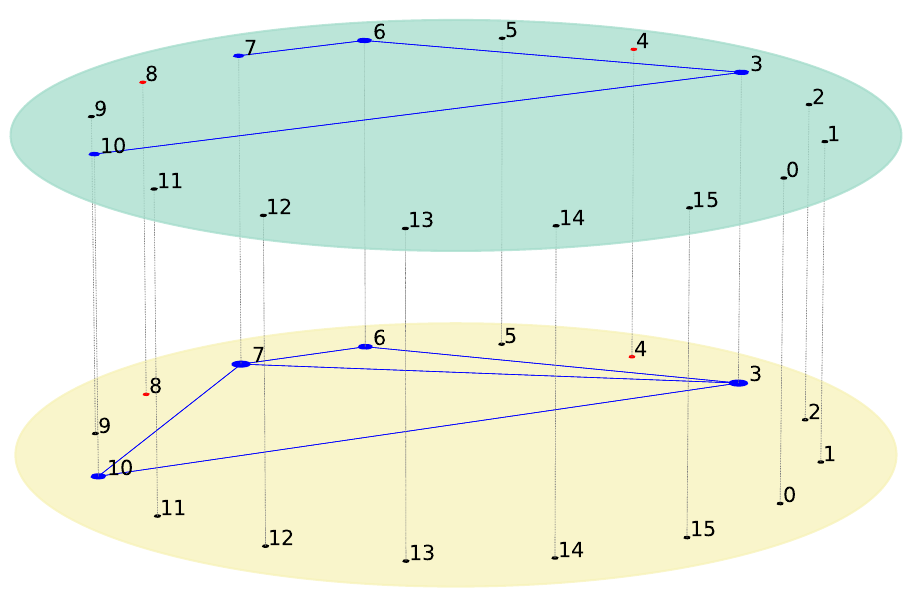}
    \subcaption{} 
    \label{Flo_family_OUR_2}
\end{subfigure}

\caption{The Padgett-Florentine Families interdependent network comprises 16 nodes. Panels \textbf{(a)}, \textbf{(b)}, and \textbf{(c)} depict the dismantling process using the HDA algorithm, while panels \textbf{(d)}, \textbf{(e)}, and \textbf{(f)} illustrate the dismantling process using the MultiDismantler algorithm. In the first step, MultiDismantler removes node 8 leading the LMCC to 6 nodes, and HDA removes node 7, reducing the LMCC to 9 nodes. Panels \textbf{(b)} and \textbf{(e)} show the network topologies after removing the most critical node identified by HDA and MultiDismantler, respectively. In the second step, the HDA algorithm removes node 8, reducing the LMCC to 5 nodes while MultiDismantler removes node 4, reducing the LMCC to 4 nodes, as shown in panels \textbf{(c)} and \textbf{(f)}. Overall, MultiDismantler demonstrates greater efficiency in reducing the LMCC compared to HDA, highlighting the clear advantage of our algorithm in identifying and removing the most critical nodes in interdependent networks.}
\label{fig-visual-family}
\end{figure}

\clearpage 
\begin{figure}[H]
    \centering

    \begin{subfigure}[b]{0.32\textwidth}
        \centering
        \includegraphics[width=\textwidth]{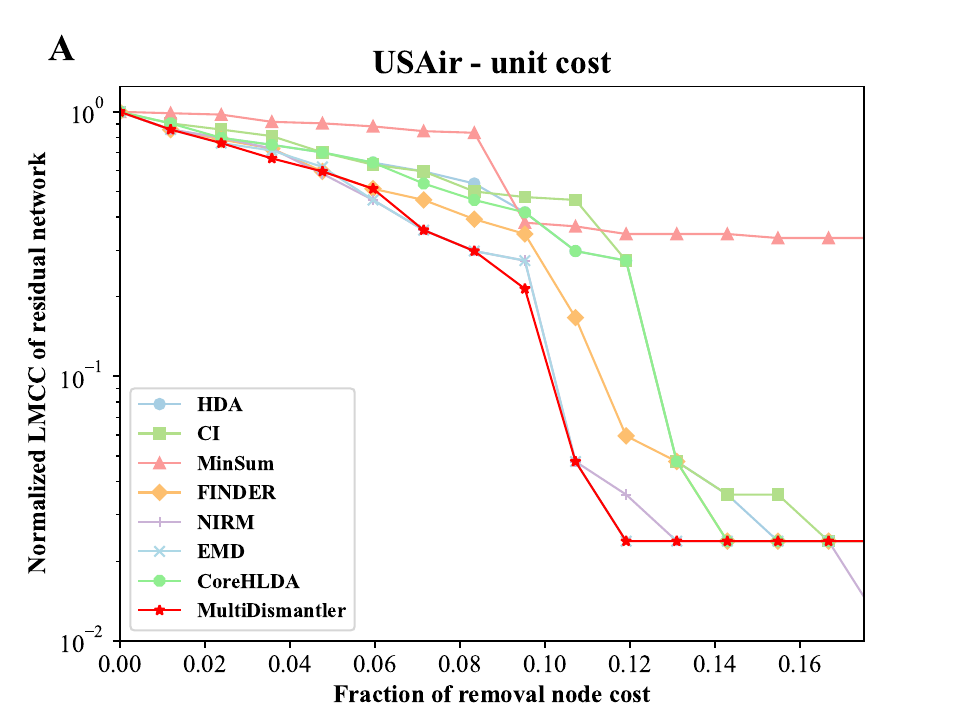}
        
        \label{fig:USAir-unit}
    \end{subfigure}
    \hspace{0.001\textwidth} 
    \hfill
    \begin{subfigure}[b]{0.32\textwidth}
        \centering
        \includegraphics[width=\textwidth]{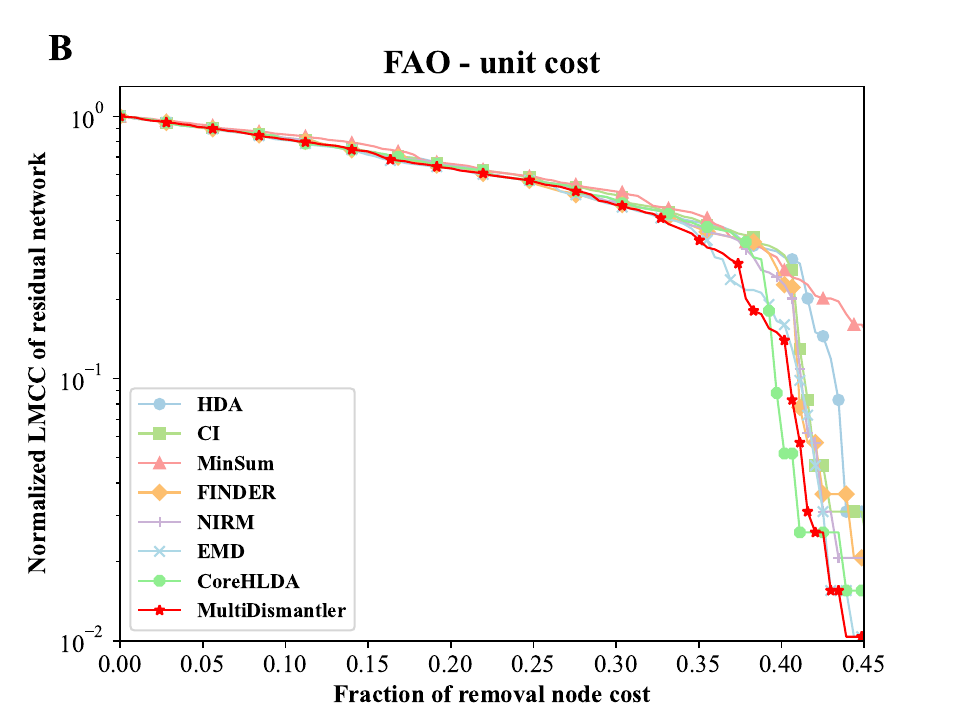}
       
        \label{FAO-unit}
    \end{subfigure}
    \hspace{0.001\textwidth} 
    \hfill
    \begin{subfigure}[b]{0.32\textwidth}
        \centering
        \includegraphics[width=\textwidth]{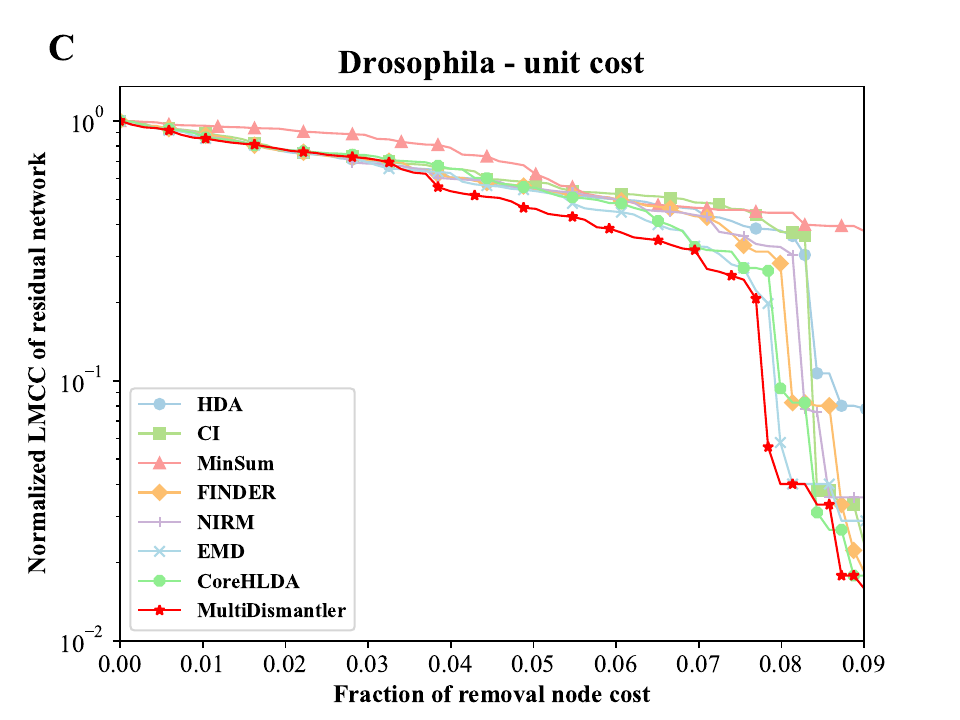}
        
        \label{Drosophila-unit}
    \end{subfigure}
    \vspace{0.5cm}

    \begin{subfigure}[b]{0.32\textwidth}
        \centering
        \includegraphics[width=\textwidth]{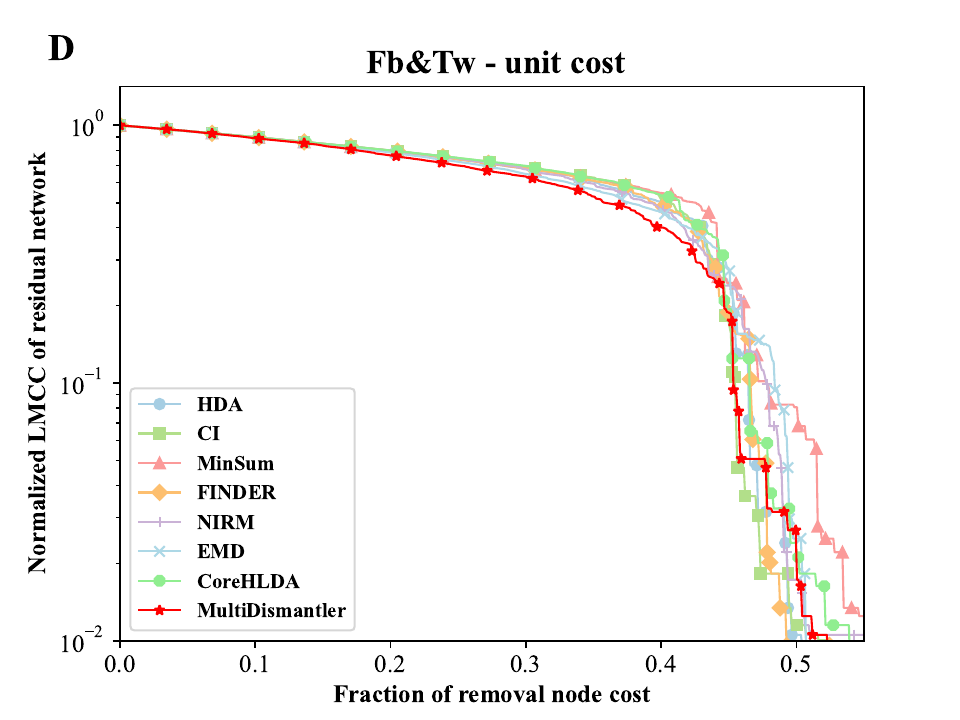}
        
        \label{Fb&Tw-unit}
    \end{subfigure}
    \hfill
    \begin{subfigure}[b]{0.32\textwidth}
        \centering
        \includegraphics[width=\textwidth]{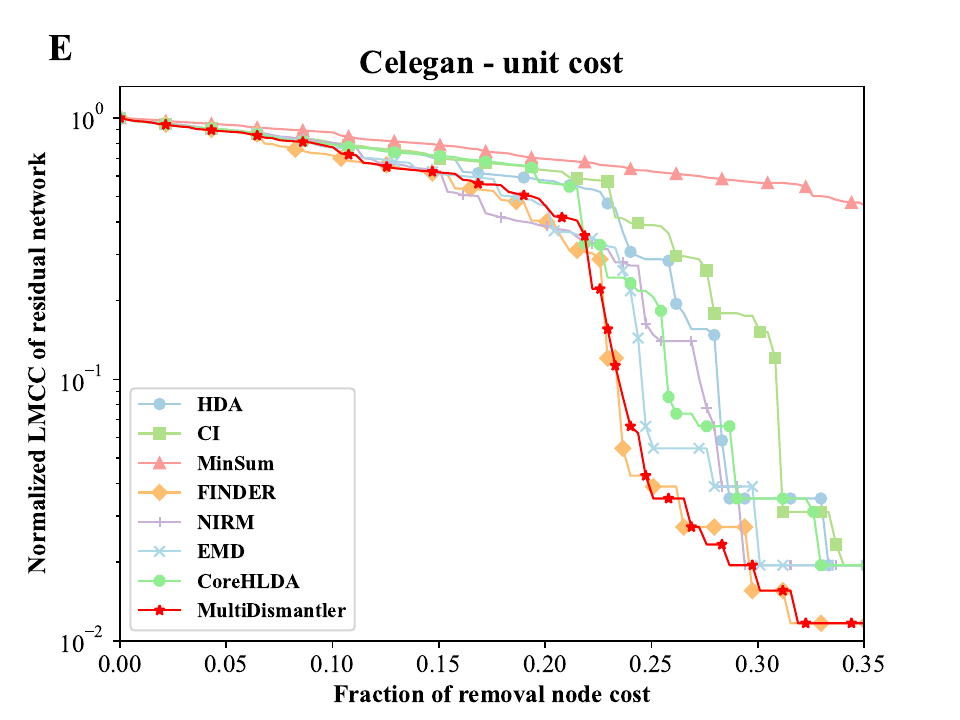}
        
        \label{Celegan-unit}
    \end{subfigure}
    \hfill
    \begin{subfigure}[b]{0.32\textwidth}
        \centering
        \includegraphics[width=\textwidth]{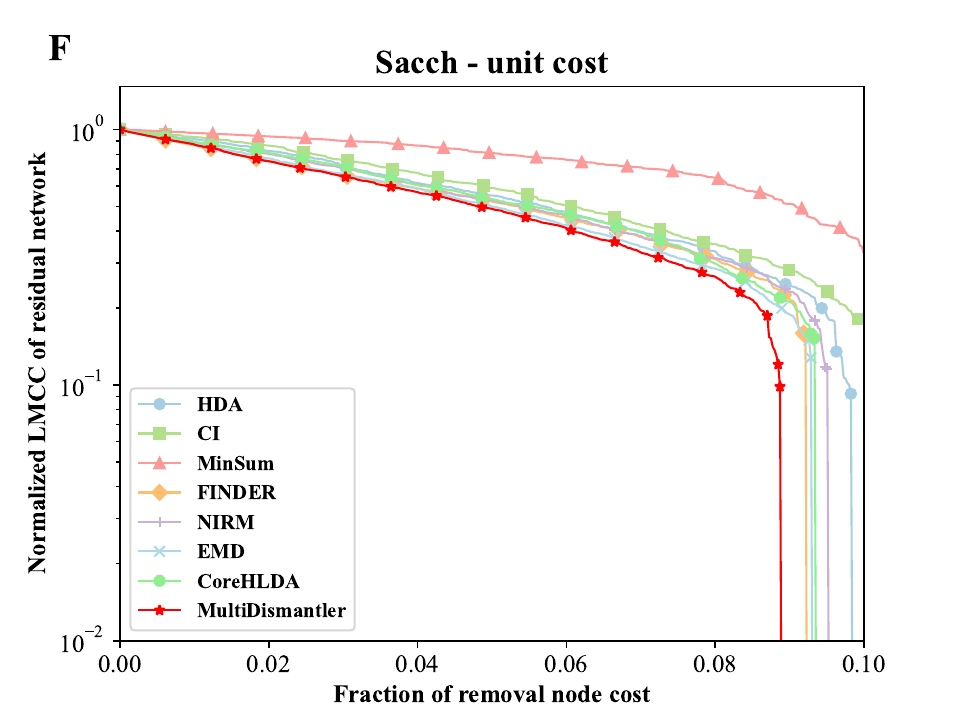}
        
        \label{Sacch-unit}
    \end{subfigure}
    \vspace{0.5cm}
    \caption{The dismantling process of real-world multiplex networks with unit costs. 
    }
    \label{fig:real-unit}
\end{figure}

\section{Full Dismantling Results of All Comparison Algorithms}

\begin{table}[H]
\caption{Dismantling performance on synthetic multiplex networks with unit costs. The best results are highlighted with bold fonts (corresponding to the lowest AUDC and $C^*$ values).}
\centering
\resizebox{\textwidth}{!}{
\begin{tabular}{c|c|ccccccccc}
    \hline
    Network size & Variable Parameters & Metric & HDA & CI & MinSum & FINDER & NIRM  & CoreHLDA&EMD& MultiDismantler \\ 
    \hline
    \multirow{6}{*}{32} & \multirow{2}{*}{g} 
    & AUDC & 0.075 & 0.079 & 0.153 & 0.081 & 0.067  & 0.076&0.067& \textbf{0.066} \\
    & & $C^*$ & 0.097 & 0.094 & 0.183 & 0.130 & 0.100  &0.152 & 0.091 & \textbf{0.089} \\
    \cline{2-11}
    \multirow{6}{*}{} & \multirow{2}{*}{$\gamma$} 
    & AUDC & 0.048 & 0.051 & 0.129 & 0.052 & 0.047  & 0.051&0.046& \textbf{0.042} \\
    & & $C^*$ & 0.066 & 0.064 & 0.155 & 0.118 & 0.070  &0.111 & 0.059 & \textbf{0.056} \\
    \cline{2-11}
    \multirow{6}{*}{} & \multirow{2}{*}{$\overline{k}$} 
    & AUDC & 0.068 & 0.074 & 0.127 & 0.071 & 0.061  & 0.067&0.059& \textbf{0.055} \\
    & & $C^*$ & 0.102 & 0.108 & 0.192 & 0.123 & 0.092  & 0.148 & \textbf{0.089}& \textbf{0.089} \\
    \hline
    \multirow{6}{*}{64} & \multirow{2}{*}{g} 
    & AUDC & 0.050 & 0.052 & 0.134 & 0.049 & 0.049  & 0.051&0.047& \textbf{0.044} \\
    & & $C^*$ & 0.088 & 0.087 & 0.198 & 0.092 & 0.085  &0.142 & 0.078 & \textbf{0.077} \\
    \cline{2-11}
    \multirow{6}{*}{}& \multirow{2}{*}{$\gamma$} 
    & AUDC & 0.029 & 0.035 & 0.088 & 0.033 & 0.030  & 0.032&0.029& \textbf{0.027} \\
    & & $C^*$ & 0.050 & 0.054 & 0.122 & 0.063 & 0.053  & 0.088 & 0.051 & \textbf{0.048} \\
    \cline{2-11}
    \multirow{6}{*}{}& \multirow{2}{*}{$\overline{k}$} 
    & AUDC & 0.055 & 0.058 & 0.137 & 0.064 & 0.051  & 0.055&0.050& \textbf{0.048} \\
    & & $C^*$ & 0.087 & 0.088 & 0.191 & 0.108 & \textbf{0.081}  &0.130 & 0.084 & \textbf{0.081} \\
    \hline
    \multirow{6}{*}{128} & \multirow{2}{*}{g} 
    & AUDC & 0.028 & 0.030 & 0.132 & 0.034 & 0.028  & 0.029&0.026& \textbf{0.025} \\
    & & $C^*$ & 0.056 & 0.059 & 0.209 & 0.067 & 0.053  &0.110 & 0.051 & \textbf{0.050} \\
    \cline{2-11}
    \multirow{6}{*}{}& \multirow{2}{*}{$\gamma$} 
    & AUDC & 0.027 & 0.029 & 0.122 & 0.025 & 0.025  & 0.027&0.023& \textbf{0.022} \\
    & & $C^*$ & 0.047 & 0.049 & 0.193 & 0.056 & 0.044  &0.070 & \textbf{0.040}& \textbf{0.040} \\
    \cline{2-11}
    & \multirow{2}{*}{$\overline{k}$} 
    & AUDC & 0.030 & 0.033 & 0.129 & 0.035 & 0.028  & 0.030&0.028& \textbf{0.026} \\
    & & $C^*$ & 0.051 & 0.055 & 0.189 & 0.063 & 0.047  &0.081 & 0.048 & \textbf{0.045} \\
    \hline
    \multirow{6}{*}{256} & \multirow{2}{*}{g} 
    & AUDC & 0.017 & 0.018 & 0.119 & 0.020 & \textbf{0.017}& \textbf{0.017}&\textbf{0.017}& \textbf{0.017} \\
    & & $C^*$ & 0.035 & 0.036 & 0.188 & 0.043 & 0.035  &0.066 & \textbf{0.034}& \textbf{0.034} \\
    \cline{2-11}
    \multirow{6}{*}{} & \multirow{2}{*}{$\gamma$} 
    & AUDC & 0.013 & 0.014 & 0.071 & 0.013 & 0.013  & 0.013&0.014& \textbf{0.012} \\
    & & $C^*$ & 0.022 & 0.022 & 0.105 & 0.030 & 0.022  &0.035 &\textbf{0.020}& \textbf{0.020} \\
    \cline{2-11}
    \multirow{6}{*}{} & \multirow{2}{*}{$\overline{k}$} 
    & AUDC & 0.021 & 0.022 & 0.117 & 0.021 & 0.020  & 0.021&0.019& \textbf{0.018} \\
    & & $C^*$ & 0.036 & 0.037 & 0.188 & 0.038 & 0.035  &0.053 & 0.035 & \textbf{0.034} \\
    \hline
    \multirow{6}{*}{512} & \multirow{2}{*}{g} 
    & AUDC & 0.019 & 0.020 & 0.115 & 0.020 & 0.018  & 0.019&0.018& \textbf{0.017} \\
    & & $C^*$ & 0.034 & 0.036 & 0.205 & 0.038 & 0.034  &0.063 & 0.034 & \textbf{0.033} \\
    \cline{2-11}
    \multirow{6}{*}{} & \multirow{2}{*}{$\gamma$} 
    & AUDC & 0.013 & 0.014 & 0.087 & 0.013 & 0.013  & 0.012&\textbf{0.011}& \textbf{0.011} \\
    & & $C^*$ & 0.020 & 0.021 & 0.139 & 0.023 & 0.019  & 0.029 & 0.019 & \textbf{0.018} \\
    \cline{2-11}
    \multirow{6}{*}{} & \multirow{2}{*}{$\overline{k}$} 
    & AUDC & 0.020 & 0.021 & 0.108 & 0.020 & 0.019  & 0.019&0.019& \textbf{0.016} \\
    & & $C^*$ & 0.031 & 0.033 & 0.173 & 0.034 & 0.029  &0.040 & 0.029 & \textbf{0.028} \\
    \hline
    \multirow{6}{*}{1024} & \multirow{2}{*}{g} 
    & AUDC & 0.014 & 0.015 & 0.106 & 0.015 & \textbf{0.013}  & 0.014&\textbf{0.013}& \textbf{0.013} \\
    & & $C^*$ & 0.027 & 0.027 & 0.173 & 0.031 & \textbf{0.026}  &0.042 & \textbf{0.026}& \textbf{0.026} \\
    \cline{2-11}
    \multirow{6}{*}{} & \multirow{2}{*}{$\gamma$} 
    & AUDC & 0.011 & 0.012 & 0.098 & 0.011 & 0.011  & 0.010&0.010& \textbf{0.009} \\
    & & $C^*$ & 0.018 & 0.018 & 0.162 & 0.019 & 0.017  & 0.022& 0.016 & \textbf{0.015} \\
    \cline{2-11}
    \multirow{6}{*}{} & \multirow{2}{*}{$\overline{k}$} 
    & AUDC & 0.016 & 0.017 & 0.098 & 0.015 & 0.015  & 0.016&0.015& \textbf{0.013} \\
    & & $C^*$ & 0.023 & 0.024 & 0.160 & 0.025 & 0.022  & 0.028 &0.021 & \textbf{0.021} \\
    \hline
\end{tabular}
}
\label{table-synthetic-unit}
\end{table}

\begin{table}[H]
\caption{Dismantling performance on synthetic multiplex networks with degree costs. The best results are highlighted with bold fonts (corresponding to the lowest AUDC and $C^*$ values).}
\centering
\resizebox{\textwidth}{!}{
\begin{tabular}{c|c|ccccccccc}
    \hline
    Network size & Variable Parameters & Metric & HDA & CI & MinSum & FINDER & NIRM  & CoreHLDA&EMD& MultiDismantler \\ 
    \hline
    \multirow{6}{*}{32} & \multirow{2}{*}{g} 
    & AUDC & 0.156 & 0.258 & 0.236 & 0.189 & 0.148  & 0.160&0.147& \textbf{0.134} \\
    & & $C^*$ & 0.209 & 0.202 & 0.295 & 0.362 & 0.224  &0.272 & 0.200 & \textbf{0.190} \\
    \cline{2-11}
    \multirow{6}{*}{} & \multirow{2}{*}{$\gamma$} 
    & AUDC & 0.112 & 0.156 & 0.208 & 0.146 & 0.110  & 0.116&0.110& \textbf{0.104} \\
    & & $C^*$ & 0.177 & 0.475 & 0.275 & 0.281 & 0.196  & 0.254 & \textbf{0.171}& 0.174\\
    \cline{2-11}
    \multirow{6}{*}{} & \multirow{2}{*}{$\overline{k}$} 
    & AUDC & 0.161 & 0.222 & 0.206 & 0.218& 0.150  & 0.156&0.144& \textbf{0.127} \\
    & & $C^*$ & 0.254 & 0.256 & 0.358 & 0.420 & 0.240  &0.319 & 0.234 & \textbf{0.224} \\
    \hline
    \multirow{6}{*}{64} & \multirow{2}{*}{g} 
    & AUDC & 0.140 & 0.167 & 0.219 & 0.192 & 0.139  & 0.140&0.135& \textbf{0.118} \\
    & & $C^*$ & 0.238 & 0.237 & 0.348 & 0.403 & 0.235  &0.318 & 0.225 & \textbf{0.208} \\
    \cline{2-11}
    \multirow{6}{*}{} & \multirow{2}{*}{$\gamma$} 
    & AUDC & 0.091 & 0.134 & 0.153 & 0.155 & 0.094  & 0.096&0.093& \textbf{0.082} \\
    & & $C^*$ & 0.177 & 0.542 & 0.252 & 0.327 & 0.192  &0.244 & 0.181 & \textbf{0.166} \\
    \cline{2-11}
    \multirow{6}{*}{} & \multirow{2}{*}{$\overline{k}$} 
    & AUDC & 0.156 & 0.192 & 0.227 & 0.217 & 0.146  & 0.153&0.147& \textbf{0.131} \\
    & & $C^*$ & 0.247 & 0.249 & 0.354 & 0.422 & 0.235  &0.310 & 0.248 & \textbf{0.230} \\
    \hline
    \multirow{6}{*}{128} & \multirow{2}{*}{g} 
    & AUDC & 0.125 & 0.140 & 0.215 & 0.139 & 0.124  & 0.126&0.120& \textbf{0.115} \\
    & & $C^*$ & 0.223 & 0.225 & 0.384 & 0.304 & 0.215  &0.297 & 0.211 & \textbf{0.208} \\
    \cline{2-11}
    \multirow{6}{*}{} & \multirow{2}{*}{$\gamma$} 
    & AUDC & 0.119 & 0.142 & 0.222 & 0.124 & 0.115  & 0.117&0.111& \textbf{0.110} \\
    & & $C^*$ & 0.209 & 0.692 & 0.377 & 0.236 & 0.205  &0.250 & \textbf{0.200}& 0.202\\
    \cline{2-11}
    \multirow{6}{*}{} & \multirow{2}{*}{$\overline{k}$} 
    & AUDC & 0.122 & 0.155 & 0.223 & 0.131 & 0.115  & 0.121&0.117& \textbf{0.105} \\
    & & $C^*$ & 0.193 & 0.200 & 0.348 & 0.276 & \textbf{0.185}  &0.244 & 0.187 & 0.187 \\
    \hline
    \multirow{6}{*}{256} & \multirow{2}{*}{g} 
    & AUDC & 0.120 & 0.128 & 0.218 & 0.126 & 0.119  & 0.118&0.119& \textbf{0.112} \\
    & & $C^*$ & 0.197 & 0.201 & 0.370 & 0.240 & 0.197&0.241 & \textbf{0.195}& 0.197\\
    \cline{2-11}
    \multirow{6}{*}{} & \multirow{2}{*}{$\gamma$} 
    & AUDC & 0.089 & 0.138 & 0.171 & \textbf{0.084} & 0.088& 0.092&0.086& 0.095 \\
    & & $C^*$ & 0.126 & \textbf{0.117} & 0.244 & 0.163 & 0.142  &0.177 & 0.124 & 0.137 \\
    \cline{2-11}
    \multirow{6}{*}{} & \multirow{2}{*}{$\overline{k}$} 
    & AUDC & 0.125 & 0.135 & 0.202 & 0.120 & 0.120  & 0.124&0.121& \textbf{0.105} \\
    & & $C^*$ & 0.196 & 0.194 & 0.355 & 0.219 & 0.192  &0.232 & 0.196 & \textbf{0.180} \\
    \hline
    \multirow{6}{*}{512} & \multirow{2}{*}{g} 
    & AUDC & 0.146 & 0.153 & 0.221 & 0.144 & 0.142  & 0.147&0.143& \textbf{0.126} \\
    & & $C^*$ & 0.210 & 0.212 & 0.401 & 0.240 & 0.208  &0.253 & 0.209 & \textbf{0.202} \\
    \cline{2-11}
    \multirow{6}{*}{} & \multirow{2}{*}{$\gamma$} 
    & AUDC & 0.095 & 0.109 & 0.168 & 0.087 & 0.092  & 0.092&0.089& \textbf{0.078} \\
    & & $C^*$ & 0.138 & 0.139 & 0.290 & 0.152 & 0.137  &0.164 & \textbf{0.134}& 0.135\\
    \cline{2-11}
    \multirow{6}{*}{} & \multirow{2}{*}{$\overline{k}$} 
    & AUDC & 0.143 & 0.151 & 0.206 & 0.131 & 0.138  & 0.138&0.136& \textbf{0.121} \\
    & & $C^*$ & 0.207 & 0.209 & 0.357 & 0.217 & 0.201  &0.225 & 0.199 & \textbf{0.197} \\
    \hline
    \multirow{6}{*}{1024} & \multirow{2}{*}{g} 
    & AUDC & 0.129 & 0.134 &0.200& 0.121& 0.127  & 0.129&0.128& \textbf{0.119} \\
    & & $C^*$ & 0.185 & 0.188 & 0.352 & 0.212 & 0.182  &0.215 & 0.184 & \textbf{0.181} \\
    \cline{2-11}
    \multirow{6}{*}{} & \multirow{2}{*}{$\gamma$} 
    & AUDC & 0.117 & 0.123 & 0.201 & 0.110 & 0.115  & 0.114&0.113& \textbf{0.108} \\
    & & $C^*$ & 0.182 & 0.184 & 0.356 & 0.188& 0.179&0.197 & \textbf{0.177}& 0.181 \\
    \cline{2-11}
    \multirow{6}{*}{} & \multirow{2}{*}{$\overline{k}$} 
    & AUDC & 0.118 & 0.126 & 0.189 & 0.112 & 0.116  & 0.118&0.116& \textbf{0.105} \\
    & & $C^*$ & 0.161 & 0.163 & 0.329 & 0.168 & 0.159  &0.177 & 0.156 & \textbf{0.155} \\
    \hline
\end{tabular}
}
\label{table-synthetic-weighted}
\end{table}

\begin{table}[H]
\caption{Dismantling performance on real-world multiplex networks with unit costs and degree costs. The best results are highlighted with bold fonts (corresponding the lowest AUDC and $C^*$ values).}
\centering

\subfloat[Unit costs]{ 
\begin{tabular}{cccccccccc} 
\hline 
Dataset & Metric & HDA & CI & MinSum & FINDER & NIRM  & CoreHLDA&EMD& MultiDismantler \\ 
\hline 
\multirow{2}{*}{UsAir} & AUDC & 0.072 & 0.077 & 0.121 & 0.061 & 0.054  & 0.071&0.054& \textbf{0.053} \\ 
& $C^*$ & 0.131 & 0.131 & 0.191 & 0.119 & \textbf{0.107}  &0.142 & \textbf{0.107}& \textbf{0.107} \\ 
\hline 
\multirow{2}{*}{FAO} & AUDC & 0.267 & 0.271 & 0.281 & 0.260 & 0.260  & 0.262 &\textbf{0.251}& 0.252\\ 
& $C^*$ & 0.439 & 0.421 & 0.481 & 0.416 & 0.416  & 0.444 & 0.421& \textbf{0.411} \\ 
\hline 
\multirow{2}{*}{Celegans} & AUDC & 0.189 & 0.203 & 0.307 & 0.167 & 0.167  & 0.183&0.168& \textbf{0.162} \\ 
& $C^*$ & 0.283 & 0.312 & 0.534 & 0.254 & 0.272  &0.315 &0.265& \textbf{0.237} \\ 
\hline 
\multirow{2}{*}{Drosophila} & AUDC & 0.065 & 0.067 & 0.084 & 0.063 & 0.063  & 0.062 &0.059& \textbf{0.056}\\ 
& $C^*$ & 0.111 & 0.102 & 0.136 & 0.106 & 0.104  &0.149 &0.099& \textbf{0.097} \\ 
\hline 
\multirow{2}{*}{Fb\&Tw} & AUDC & 0.337 & 0.339 & 0.347 & 0.336 & 0.336  & 0.342 &0.333& \textbf{0.319} \\ 
& $C^*$ & 0.472 & \textbf{0.471} & 0.516 & 0.478 & 0.490  &0.502 &0.500& 0.478 \\ 
\hline 
\multirow{2}{*}{Netsci} & AUDC & 0.051 & 0.055 & 0.118 & 0.049 & \textbf{0.049} & 0.050 &0.052& 0.051 \\ 
& $C^*$ & \textbf{0.084} & 0.102 & 0.185 & 0.089 & 0.094  &0.274 &0.114 & 0.095 \\ 
\hline 
\multirow{2}{*}{Sacchpomb} & AUDC & 0.056 & 0.060 & 0.079 & 0.052 & 0.054  & 0.054 &0.051& \textbf{0.049} \\ 
& $C^*$ & 0.099 & 0.102 & 0.105 & 0.092 & 0.095  &0.096 &0.093& \textbf{0.089} \\ 
\hline 
\multirow{2}{*}{Homo} & AUDC & 0.030 & 0.032 & 0.054 & 0.030 & 0.030  & 0.030&0.028& \textbf{0.027} \\ 
& $C^*$ & 0.065 & 0.066 & 0.100 & 0.064 & 0.064  &0.071 &0.063& \textbf{0.060} \\ 
\hline 
\multirow{2}{*}{Sanremo2016} & AUDC & 0.002 & 0.002 & 0.004 & \textbf{0.001} & 0.002  & 0.002&\textbf{0.001}& \textbf{0.001} \\ 
& $C^*$ & 0.004 & \textbf{0.003} & 0.023 & 0.008 & 0.008  &0.014 &0.008 & 0.004 \\ 
\hline 
\end{tabular} 
\label{table-real-world-unit}
}
\vspace*{0.02\linewidth}

\subfloat[Degree costs]{ 
\begin{tabular}{cccccccccc} 
\hline 
Dataset & Metric & HDA & CI & MinSum & FINDER & NIRM  & CoreHLDA&EMD& MultiDismantler \\ 
\hline 
\multirow{2}{*}{UsAir} & AUDC & 0.264 & 0.276 & 0.298 & 0.248 & 0.220  & 0.263&0.221& \textbf{0.215} \\ 
& $C^*$ & 0.441 & 0.441 & 0.457 & 0.502 & 0.376  & 0.456& 0.358 & \textbf{0.349} \\ 
\hline 
\multirow{2}{*}{FAO} & AUDC & 0.555 & 0.558 & 0.560 & 0.539 & 0.549  & 0.552&0.539& \textbf{0.520} \\ 
& $C^*$ & 0.804 & 0.787 & 0.828 & 0.862 & \textbf{0.781}  &0.808 & 0.785 & 0.804 \\ 
\hline 
\multirow{2}{*}{Celegans} & AUDC & 0.388 & 0.407 & 0.419 & 0.407 & 0.362  & 0.380&0.360& \textbf{0.315} \\ 
& $C^*$ & 0.514 & 0.547 & 0.684 & 0.784 & 0.502  &0.540 & 0.494 & \textbf{0.480} \\ 
\hline 
\multirow{2}{*}{Drosophila} & AUDC & 0.287 & 0.293 & 0.302 & 0.301 & 0.284  & 0.284&0.275& \textbf{0.268} \\ 
& $C^*$ & 0.412 & 0.389& 0.421 & 0.544 & 0.402  & 0.453 & \textbf{0.385}& 0.399 \\ 
\hline 
\multirow{2}{*}{Fb\&Tw} & AUDC & 0.483 & 0.487 & 0.461 & \textbf{0.410} & 0.482  & 0.487&0.469& 0.417 \\ 
& $C^*$ & 0.628 & 0.628 & 0.664 & 0.864 & 0.642  &0.653 & 0.644 & \textbf{0.620} \\ 
\hline 
\multirow{2}{*}{Netsci} & AUDC & 0.219 & 0.228 & 0.267 & 0.302 & 0.212  & 0.215&0.215& \textbf{0.187} \\ 
& $C^*$ & \textbf{0.310} & 0.339 & 0.427 & 0.661 & 0.328  & 0.516& 0.355 & 0.344 \\ 
\hline 
\multirow{2}{*}{Sacchpomb} & AUDC & 0.404 & 0.418 & 0.396 & 0.356 & 0.396  & 0.399&0.382& \textbf{0.310} \\ 
& $C^*$ & 0.565 & 0.574 & 0.539& 0.596 & 0.559  & \textbf{0.516}& 0.553 & 0.658 \\ 
\hline \multirow{2}{*}{Homo} & AUDC & 0.315 & 0.323 & 0.357 & 0.305 & 0.317  & 0.317&0.305& \textbf{0.291} \\ 
& $C^*$ & 0.510 & 0.514 & 0.534 & 0.614 & 0.510  &0.520 & 0.501& \textbf{0.495} \\ 
\hline \multirow{2}{*}{Sanremo2016} & AUDC & 0.164 & 0.184 & 0.168 & \textbf{0.149} & 0.178  & 0.163&0.152& 0.150 \\ 
& $C^*$ & 0.348 & 0.362 & 0.355 & 0.350& 0.326  &0.355 &0.327 & \textbf{0.321} \\ 
\hline 
\end{tabular} 
\label{table-real-world-weighted}
}
\label{table-real-world}
\end{table}





\section{MGNN Framework}
\begin{figure}[H]  
\centering  
\includegraphics[width=\linewidth]{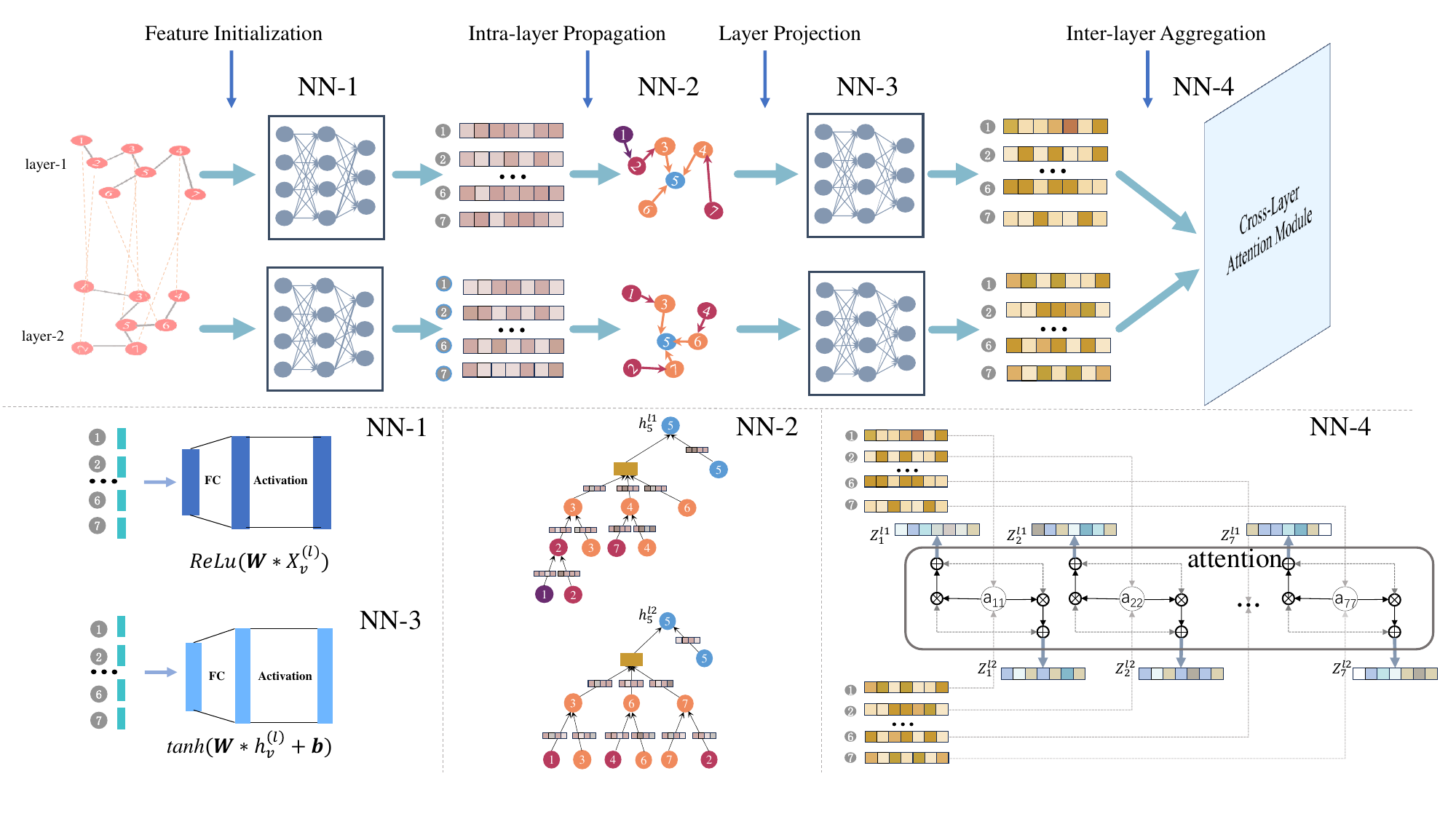}  
\caption{The overall framework of the MGNN, which includes feature initialization, intra-layer message propagation, layer projection, and inter-layer information aggregation. The intra-layer process performs graph convolution on each layer to update node embeddings based on their local neighborhood structure. The inter-layer process aggregates node embeddings across different layers using the attention mechanism, enabling the model to capture intra- and inter-layer structure information.}  
\label{fig:encoder}  
\end{figure}

\bibliography{sample}